\documentclass[preprint,3p,times,onecolumn]{elsarticle}

\usepackage{amssymb,graphicx}
\usepackage{amsmath,bm}

\newcommand{\bea}{\begin{eqnarray}}
\newcommand{\eea}{\end{eqnarray}}
\newcommand{\bes}{\begin{subequations}}
\newcommand{\ees}{\end{subequations}}

\begin{document}

\begin{frontmatter}
\title{Non-autonomous bright matter wave solitons in spinor Bose-Einstein Condensates}

\author{T. Kanna\corref{tk}}\ead{kanna\_phy@bhc.edu.in}
\author{R. Babu Mareeswaran}
\author{K. Sakkaravarthi}\ead{ksakkaravarthi@gmail.com}
\address{Post Graduate and Research Department of Physics, Bishop Heber College, Tiruchirapalli--620 017, Tamil Nadu, India}
\cortext[tk]{Corresponding author}

\begin{abstract}
We investigate the dynamics of bright matter wave solitons in spin-1 Bose-Einstein condensates with time modulated nonlinearities. We obtain soliton solutions of an integrable autonomous three-coupled Gross-Pitaevskii (3-GP) equations using Hirota's method involving a non-standard bilinearization. The similarity transformations are developed to construct the soliton solutions of non-autonomous 3-GP system. The non-autonomous solitons admit different density profiles. An interesting phenomenon of soliton compression is identified for kink-like nonlinearity coefficient with Hermite-Gaussian-like potential strength. Our study shows that these non-autonomous solitons undergo non-trivial collisions involving condensate switching.
\end{abstract}

\begin{keyword}
Spinor Bose-Einstein Condensate \sep three-coupled Gross-Pitaevskii equation \sep similarity transformation \sep Hirota's bilinearization method \sep bright soliton solution \sep soliton interaction \newline ------------------------------------------------------------------------------------------------------------------------------------------- \newline Journal Reference: {\it Physics Letters A}~~~~~~~~doi: 10.1016/j.physleta.2013.11.002
\end{keyword}
\end{frontmatter}

\section{Introduction}
The study on Bose-Einstein condensates (BECs) has envisaged tremendous growth since from their experimental realizations \cite{becexp1,becexp2,becexp3}. Particularly, the study of multicomponent superfluid system has been a tantalizing goal of low-temperature physics for the past few decades \cite{Ho}. One of the recent developments in BECs is the study of dilute Bose gases with internal degrees of freedom \cite{ohmi}. The experimental demonstrations of spinor BECs ${}^{23}${Na} \cite{Stamper} and ${}^{87}${Rb} \cite{Matthews} in optical traps have paved way to study the non-trivial properties of multicomponent spinor condensates. Under the magnetic potential traps the spin degree of freedom is frozen whereas the optical potential trap enables the spin degree of freedom to be free \cite{ho1}.

The multicomponent solitons in spinor condensates with hyperfine spin $F=1$ have been investigated in detail and different types of solitons, namely, bright solitons \cite{Ieda}, dark solitons \cite{Uchi}, gap solitons \cite{Kivshar}, and bright-dark soliton complexes \cite{Frantz} have been reported. Spinor condensates with hyperfine spin $F$ =2, have also attracted considerable attention \cite{ger}. From a theoretical perspective, the evolution of spinor condensates is described by a set of multiple-coupled Gross-Pitaevskii (GP) equations, in the mean field approximation. This set of multicomponent GP equations is a classical nonlinear evolution equation (with the nonlinearity originating from the interatomic interactions) and, as such, it permits the study of variety of interesting purely nonlinear phenomena. These multicomponent GP equations have primarily been studied by treating the condensate as a purely nonlinear coherent matter wave, i.e., from the viewpoint of the dynamics of nonlinear waves. This mean-field approximation is valid for large values of $N$, where $N$ is the number of particles \cite{Pethick}. Then the energy will be proportional to $N^2$ and one can consider the field operators as complex numbers and obtain the multiple-coupled GP equations as the evolution equation for spinor condensates. However, one can also develop quantum many body treatment for the spinor condensates which is beyond the scope of the present work. A comparison between the mean-field theory and many-body treatment for single component BEC and spin-1 BECs can be found in Refs.\cite{Str} and Ref.\cite{weib}, respectively.

The multiple-coupled GP system obtained by mean-field theory has analogy with multicomponent coherently coupled nonlinear Schr\"odinger system arising in the context of nonlinear optics \cite{kivbook,tkjpa}. In references \cite{Ieda,Uchi}, various soliton solutions of a set of integrable autonomous 3-coupled GP equations with constant attractive and repulsive nonlinearity coefficients have been reported by applying the sophisticated Inverse Scattering Transform (IST) method. The soliton solutions of spinor condensates can be classified as polar (or anti-ferromagnetic) and ferromagnetic solitons depending upon their total spin \cite{ho1,Ieda}. The former arises for zero spin while the latter results for non-zero spin.

An exiting possibility in spinor condensates is that the spin-exchange interaction can be tuned by employing optical (as well as magnetic) \cite{{Theis},{Jack}} and microwave \cite{micro} fields. The Feshbach resonance mechanism has provided good tool for studying the dynamics of spinor BECs with spatially/temporally varying nonlinearities. Particularly, the two $s$-wave scattering lengths ($a_0$ and $a_2$) in spinor condensates can be tuned for wide range of values using optical field \cite{Theis}. These theoretical and experimental studies prompted us to investigate the evolution of spinor condensates in the presence of time-varying nonlinearities and external optical/magnetic traps.

The motivation of present study arises from another fact that certain non-autonomous coupled nonlinear evolution equations can be transformed to a set of integrable autonomous equations by means of special similarity transformations. Here, non-autonomous refers to explicit appearance of time in the nonlinear evolution equation. Otherwise, the evolution equation is said to be autonomous (time acts only as an independent variable). This type of transformation was first proposed by Serkin {\it et al} \cite{similarity} for the non-autonomous nonlinear Schr\"odinger (NLS) equation with external potential and several solvable choices of nonlinearities and potentials were reported. Subsequently, the study has been extended to two-coupled NLS equations \cite{huss} and their $N$-component version in Ref.\cite{raj}. Particularly, in Ref.\cite{raj}, the non-autonomous soliton solutions were constructed with the knowledge of available bright \cite{tkopt,tkpra}, dark \cite{ohtads,kivds} and bright-dark \cite{kivds,mv} soliton solutions of the integrable multiple-coupled NLS type equations. This kind of transformation of non-autonomous system to an integrable autonomous system is possible mainly due to the explicit dependence of nonlinearity coefficients on time and due to the presence of external potential.

In the present work, the focus is on the study of dynamics of spinor BECs under time-varying spin-exchange interaction and inhomogeneous external potential. For this purpose, we consider a dilute gas of optically trapped bosonic atoms for a spinor BEC, with hyperfine spin $F = 1$, in the presence of external potential with time-varying mean-field and spin-exchange interactions. The evolution of such spinor condensates can be described by the following set of three-coupled GP (3-GP) equations which is a generalization of the 3-GP system given in Refs. \cite{Ieda,ragav}.
\bes\bea
i\psi_{\pm1,t}& = &-\psi_{\pm1,xx}+(c_0(t)+c_2(t))(|\psi_{\pm1}|^2+2|\psi_{0}|^2)\psi_{\pm1}+2c_2(t)\psi_0^2\psi_{\mp1}^* +(c_0(t)-c_2(t))|\psi_{\mp1}|^2\psi_{\pm1}+V_{ext}(x,t)\psi_{\pm1},~~\\
i\psi_{0,t}& =&-\psi_{0,xx}+(c_0(t)+c_2(t))(|\psi_{+1}|^2+|\psi_{-1}|^2)\psi_{0}+2c_0(t)|\psi_0|^2\psi_{0} +2c_2(t)\psi_{-1}\psi_{0}^*\psi_{+1}+V_{ext}(x,t)\psi_{0}.
\eea\ees
In Eq.(1), $\psi_{+1},\psi_{0},\psi_{-1}$ are respectively the wave functions of the three spin components, with magnetic spin quantum numbers $m_F=+1,0,-1$. The one-dimensional coupling coefficients $c_0(t)$ and $c_2(t)$ denote the mean-field and spin-exchange interactions, respectively and are given by $c_0(t)$ = $\frac{4\pi\hbar^2}{3m}(a_0(t) + 2a_2(t))$ and $c_2(t)$ = $\frac{4\pi\hbar^2}{3m}(a_2(t) - a_0(t))$. Here the $s$-wave scattering lengths $a_0(t)$ and $a_2(t)$ are tuned by optical means \cite{Theis}, which in turn makes the mean-field and spin-exchange interactions as time-dependent functions \cite{Wen}, which are treated as constants in the earlier works \cite{Ieda}. Particularly, for the integrable choice $c_0 = c_2 = -c~(+c)$, where $c$ is a positive real constant, corresponding to repulsive (attractive) condensates \cite{tkpain}, bright (dark) soliton solutions of Eqs.  (1) are obtained in the absence of external potential $V_{ext}$ in Ref.\cite{Ieda} (Ref. \cite{Uchi}). There also exists another integrable choice $c_2=0 $ with constant $c_0$, for which Eqs.  (1) reduces to the integrable 3-component Manakov type equations and admits bright (dark) solitons for $c_0 < 0$ ($c_0>0$) \cite{{tkopt},{tkpra},{ohtads},{kivds},{mv}}. The total number of atoms $N_T$, total spin $F_T$ and momentum $P_T$ are given by $N_{T} = \int dx\bf{\Psi}^{\dag}\bf{\Psi}$, $F_{T} = \int dx \bf{\Psi}^{\dag} \cdot \textbf{\emph{f}} \cdot \bf{\Psi}$ and $P_T = -i\hbar \int dx \bf{\Psi}^{\dag}\cdot \partial_x\bf{\Psi}$, respectively, where ${\bf \Psi}=(\psi_{+1},~\psi_0,~\psi_{-1})^{\textit{T}}$ and $\textbf{\emph{f}}=(\emph{f}^{~x},\emph{f}^{~y},\emph{f}^{~z})^{\textit{T}}$ in which $\emph{f}^{~i}$'s are the three $3\times 3$ spin-1 matrices \cite{Ieda}.

Now, we are interested in studying the dynamics of spinor solitons in the above non-autonomous 3-GP system (1) for the choice $c_0(t) = c_2(t) = -c(t)$. The corresponding non-autonomous 3-GP system is
\bes\bea
\hspace{-2cm}&& i\psi_{\pm1,t}=-\psi_{\pm1,xx}-2c(t)(|\psi_{\pm1}|^2+2|\psi_{0}|^2)\psi_{\pm1}-2c(t)\psi_0^2\psi_{\mp1}^* +V_{ext}(x,t)\psi_{\pm1},\\
\hspace{-2cm}&&i\psi_{0,t}=-\psi_{0,xx}-2c(t)(|\psi_{+1}|^2+|\psi_0|^2+|\psi_{-1}|^2)\psi_{0} -2c(t)\psi_{-1}\psi_{0}^*\psi_{+1}+V_{ext}(x,t)\psi_{0},~
\eea\label{ncgp}\ees
where $t$ and $x$ are time and spatial coordinates. Generally, the time- and/or space- modulated external confining potential $V_{ext}$ can be chosen in the form of harmonic, double-well or optical lattice potential \cite{similarity}. In our case, we choose the one-dimensional harmonic external potential $V_{ext}$ which is same for all the three spin components as $V_{ext}=(1/2)\Omega^2(t)x^2$, where $\Omega(t)$ is the strength of the potential.

The rest of the paper is organized in the following manner. In Sec. \ref{trans}, we present the similarity transformation which transforms the non-autonomous 3-GP equations (\ref{ncgp}) into a set of autonomous 3-GP equations with an integrable condition. We construct the bright one- and two- matter wave soliton solutions of autonomous 3-GP equations by using the Hirota's method in Sec. \ref{autosol}. In Sec. \ref{nonautosol}, we obtain the explicit soliton solutions of  non-autonomous 3-GP equations by inverting the similarity transformations and analyse the nature of time-varying nonlinear effects by considering the kink-like nonlinearity as an example. In Sec. \ref{collisions}, we explore the different types of non-autonomous matter wave soliton interactions and the results will be compared with autonomous matter wave solitons. The results are summarized in final section.

\section{Similarity Transformation }\label{trans}
The first step of our study on Eqs.  (\ref{ncgp}) is to look for a similarity transformation that transforms the non-autonomous spinor 3-GP system (\ref{ncgp}) to a standard integrable 3-GP system. This will be of use in identifying the explicit form of time-dependent nonlinearity coefficient and the nature of corresponding confining potential which can support soliton/soliton-like structures in spinor BECs.\\

\indent  We apply the following similarity transformation
\bes\bea
&&(\psi_{+1},~\psi_{0},~\psi_{-1})^T = \xi_1\sqrt{{c}(t)}~e^{i\tilde{\theta}(x,t)} (q_1,~q_2,~q_3)^T,\quad
\eea
where
\bea
&&\tilde{\theta}(x,t) = \left[-\frac{d}{dt}(\ln c)\right]\frac{x^2}{2} + 2 \xi_1^2 \xi_2 \left({c}x - 2 \xi_2 \xi_1^2\int_0^t {c}^2 dt\right),\\
&&X = ~\xi_1 \left[{c} x - 2\sqrt {2}\xi_2 \xi_1^2\int_0^t {c}^2 dt\right],\\
&&T = \xi_1^2 \int_0^t {c}^2 dt,
\eea\label{str}\ees
and $\xi_j,~j=1,2,$ are arbitrary real constants, to Eqs.  (\ref{ncgp}). The resulting transformed equations reduce to the following set of known integrable three-component GP equations \cite{Ieda,tkpain}
\bes\bea
&iq_{1,T}+ q_{1,XX}+2(|q_1|^2+2|q_2|^2)q_1 + 2q_2^2q_3^*=0,\\
&iq_{2,T}+ q_{2,XX}+2(|q_1|^2+|q_2|^2+|q_3|^2)q_2+ 2q_1q_3q_2^*=0,\\
&iq_{3,T}+ q_{3,XX}+2(2|q_2|^2+|q_3|^2)q_3 + 2q_2^2q_1^*=0,
\eea\label{ccnls}
with a constraint which can be expressed in the form of Riccati equation
\bea
\frac{dy}{dt}-y^2 - \Omega^2(t)=0,
\label{Riccati}
\eea
\ees
where $y~ =c_t/c$ is the dependent variable.
The aim of the present study is to construct autonomous soliton solutions of (\ref{ccnls}) using Hirota's approach and then make use of the solutions to explore the interesting dynamics of non-autonomous matter wave solitons in the non-autonomous spinor BEC system (\ref{ncgp}).

\section{Bright matter wave soliton solutions of 3-GP system (\ref{ccnls})} \label{autosol}
The explicit soliton solutions of the integrable 3-GP equations (\ref{ccnls}) have been obtained in Ref. \cite{Ieda} by applying the sophisticated IST method. In this section, we construct those exact bright matter wave soliton solutions of Eqs.  (\ref{ccnls}) by using the Hirota's bilinearization procedure \cite{hirota}. These explicit solutions and their subsequent analysis presented briefly in the following sub-sections are necessary to construct the non-autonomous matter wave solitons of system (\ref{ncgp}) and to gain insight into their interesting collision dynamics, as mentioned in the introduction. During the bilinearization procedure, to deal with the spin-mixing nonlinearities, we have to introduce an auxiliary function to obtain consistent general soliton solutions of system (\ref{ccnls}), which is an uncommon practice for bilinearizing nonlinear evolution equations. The interested readers can refer to Refs. \cite{tkjpa,gil} for further details. A brief discussion on the one- and two- bright matter wave soliton solutions of Eqs.  (\ref{ccnls}) is given in the following sub-sections.

\subsection{Bright matter wave one-soliton solution}
The bilinear equations of system (\ref{ccnls}) obtained by performing a rational transformation $q_j=\frac{g^{(j)}}{f}$, $j=1,2,3$, to Eqs.  (\ref{ccnls}) with the introduction of an auxiliary function `$s$' are
\bes\bea
&&(iD_T+D_X^2) g^{(j)} \cdot f = (-1)^{(j+1)} s \cdot g^{(4-j)*}, \quad j=1,2,3,\\
&&D_X^2 f \cdot f = 2 \left(|g^{(1)}|^2+2|g^{(2)}|^2+|g^{(3)}|^2\right),\\
&&s \cdot f = g^{(1)}\cdot g^{(3)}-(g^{(2)})^2,
\eea\label{beq}\ees
where $g^{(j)}$'s and $s$ are complex functions while $f$ is a real function to be determined. Here, the Hirota's bilinear operators $D_T$ and $D_X$ \cite {hirota} are defined as below:
\bea
\hspace{-1cm}D_X^{p}D_T^{q}(a\cdot b) =\bigg(\frac{\partial}{\partial X}-\frac{\partial}{\partial X'}\bigg)^p\bigg(\frac{\partial}{\partial T}-\frac{\partial}{\partial T'}\bigg)^q a(X,T)b(X',T')\Big|_{(X=X',T=T')}.\nonumber
\eea
The bright soliton solutions can be obtained by carrying out the standard steps of Hirota's method \cite{tkjpa,hirota} and the results are given below.

\indent The general bright one-soliton solution of autonomous 3-GP Eqs.  (\ref{ccnls}) can be written as
\bes\bea
&&q_{j}(X,T) = \frac{\alpha_1^{(j)} e^{\eta_1}+e^{2\eta_1+\eta_1^*+\delta_{11}^{(j)}}}{1+e^{\eta_1+\eta_1^*+R_1}+e^{2\eta_1+2\eta_1^*+\epsilon_{11}}},\quad j=1,2,3,
\eea
where
\bea
&&\eta_1 = k_1(X+ik_1T),\\
&&e^{R_1}=\frac{{(|\alpha_1^{(1)}|^2+2|\alpha_1^{(2)}|^2+|\alpha_1^{(3)}|^2)}}{(k_1+k_1^*)^2}, \quad e^{\epsilon_{11}}=\frac{|\Gamma_1|^2}{(k_1+k_1^*)^4}, \\
&& e^{\delta_{11}^{(j)}}=\frac{(-1)^{j+1} \alpha_1^{(4-j)*} \Gamma_1}{(k_1+k_1^*)^2}, \quad j=1,2,3,\quad \Gamma_1=\alpha_1^{(1)}\alpha_1^{(3)}-(\alpha_1^{(2)})^2.
\eea\label{s1s}
\ees
The auxiliary function `$s$' is determined as $s=\Gamma_1 e^{2\eta_1}$.

The above bright matter wave soliton solution of system (\ref{ccnls}) can be classified as ferromagnetic soliton (FS) and polar soliton (PS) based on the total spin following Ref. \cite{Ieda}. When the coefficient $\Gamma_1$ in the expression for the auxiliary function becomes zero (non-zero) it can be verified that the total spin becomes non-zero (zero), as defined in the introduction following Refs. \cite{ho1,Ieda}. Thus, for $\Gamma_1=0$ ($\Gamma_1 \neq 0$) one can have ferromagnetic soliton (polar soliton). This clearly shows that the spin-mixing nonlinearity determines the nature of soliton whether it is FS or PS \cite{Ho}. Here, we give the explicit forms and briefly discuss the dynamics of both FS and PS of system (\ref{ccnls}), though they are studied in detail in Ref. \cite{Ieda}, for completeness and for getting further impetus into the dynamics of non-autonomous solitons of system (\ref{ncgp}). Also, this revisit on the autonomous FS and PS solitons has lead us to identify the parametric choice for which the FSs can also undergo elastic collision, as will be seen in section \ref{collisions}.

\begin{figure}[h]
\centering\includegraphics[width=0.55\linewidth]{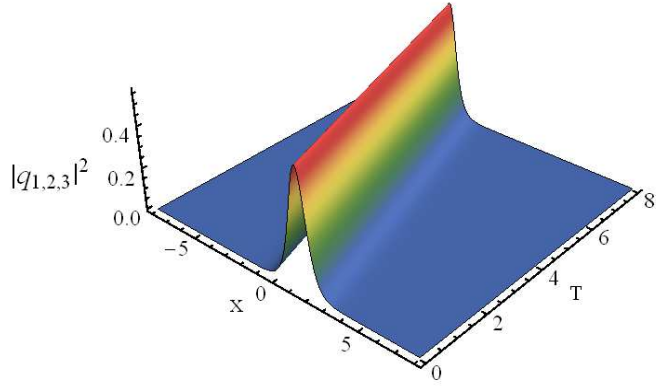}
\caption{Propagation of ferromagnetic bright matter wave soliton.}
\label{fs1sol}
\end{figure}

\noindent\underline{Case(i): Ferromagnetic solitons}\\
\indent The total spin of the FSs is non-zero which results for $\Gamma_1=0$ and for this choice the auxiliary function `$s$' vanishes. This type of solitons have the standard ``sech" profile and the corresponding expression can be rewritten from the general soliton solution (\ref{s1s}) as
\bea
q_j&=& A_j~\mbox{sech}\left(\eta_{1R}+R_1/2\right)e^{i\eta_{1I}}, \quad j=1,2,3,
\label{ms}
\eea
where $A_j$ is the amplitude of FS in the $j^{th}$ component and is defined as $A_j=\frac{\alpha_1^{(j)}}{2}e^{\frac{-R_1}{2}} \equiv \frac{k_{1R}{\alpha_1^{(j)}}}{\sqrt{|\alpha_1^{(1)}|^2+2|\alpha_1^{(2)}|^2+|\alpha_1^{(3)}|^2}},~j=1,2,3$, $R_1=\mbox{ln}\left[\frac{1}{4 k_{1R}^2} \left(|\alpha_1^{(1)}|^2+2|\alpha_1^{(2)}|^2+|\alpha_1^{(3)}|^2\right)\right]$,
$\eta_{1R}=k_{1R}(X-2k_{1I}T)$ and $\eta_{1I}=k_{1I}{X}+(k_{1R}^2-k_{1I}^2){T}$. Note that the quantity $\frac{{\alpha_1^{(j)}}}{\sqrt{|\alpha_1^{(1)}|^2+2|\alpha_1^{(2)}|^2+|\alpha_1^{(3)}|^2}}$ represents the spin polarization. This shows that the $\alpha_i^{(j)}$-parameters determine the spin polarization and play an important role in the dynamics of matter wave solitons. The ferromagnetic soliton (\ref{ms}) is characterized by three arbitrary complex parameters. The propagation of FS in the system (\ref{ccnls}) is shown in Fig.~\ref{fs1sol} for the parameters $k_1 = 1.5 - 0.3i$ and $\alpha_1^{(1)} = \alpha_1^{(2)} = \alpha_1^{(3)} = 0.2$.

\begin{figure}[h]
\centering
\includegraphics[width=0.4\linewidth]{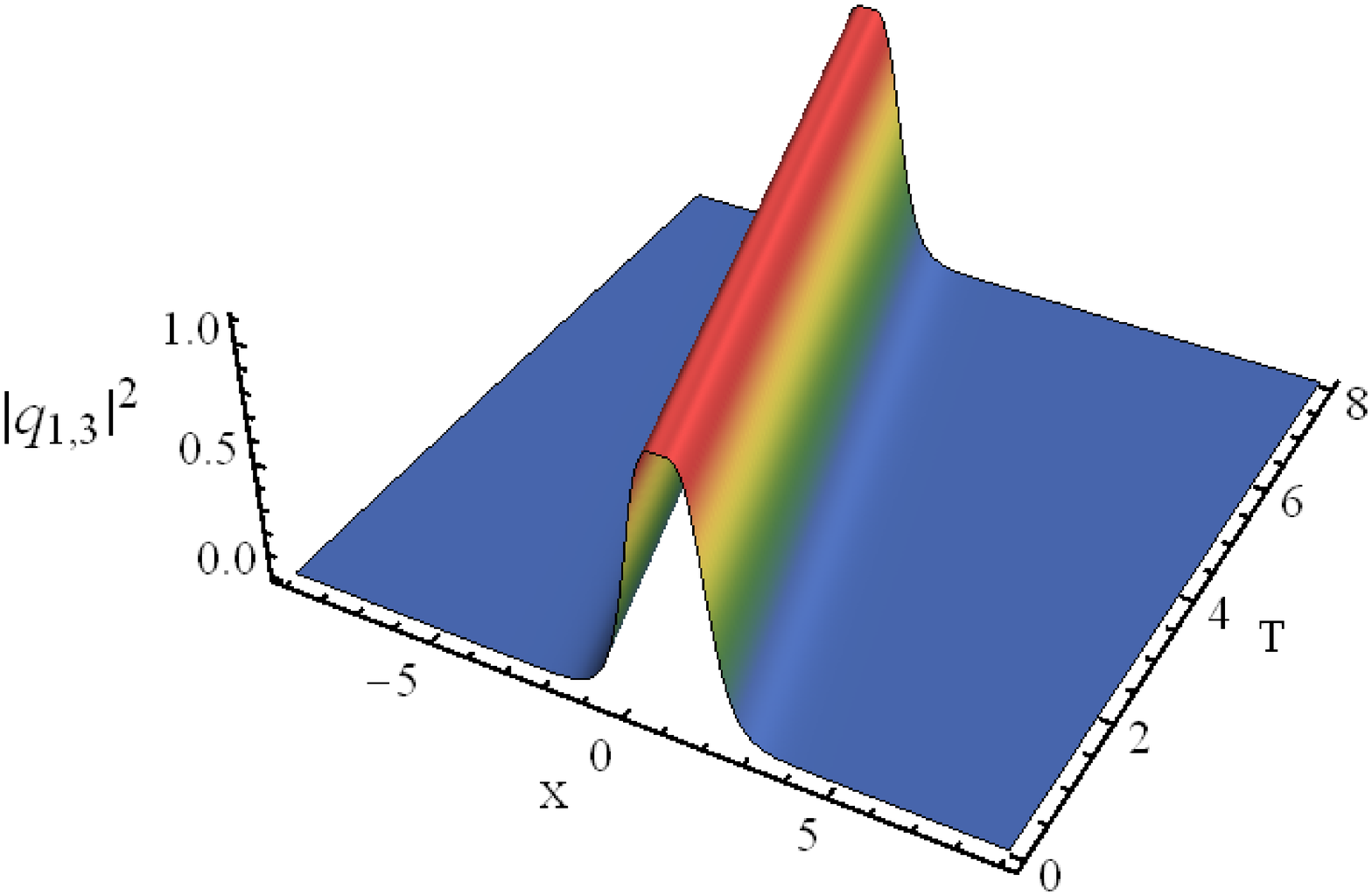}~~~~~~~~\includegraphics[width=0.4\linewidth]{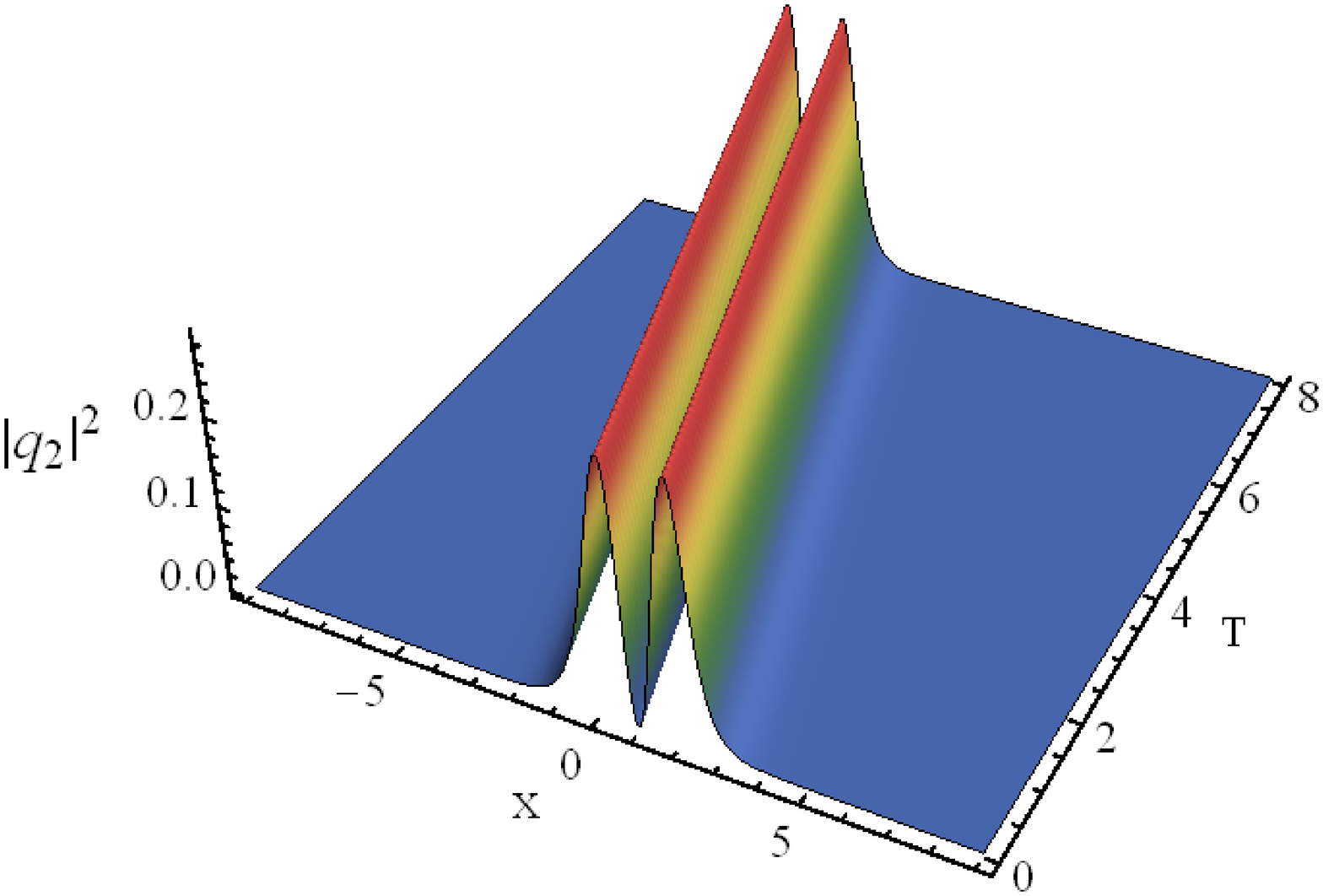}
\caption{Propagation of polar soliton with flat-top (double-hump) profile in $q_1$ and $q_3$ $(q_2)$ components in the integrable 3-GP system (\ref{ccnls})}.
\label{genpolar}
\end{figure}

\noindent\underline{Case(ii): Polar (anti-ferromagnetic) solitons}\\
\indent The polar solitons having zero total spin result for the general choice, $\Gamma_1 \neq 0$, and hence the auxiliary function `$s$' becomes non-zero. The expression for such PSs can be written in a compact form as
\bea
\hspace{-1.5cm}q_j=\frac{2A_j\left[\mbox{cos}(P_j)\mbox{cosh}\left(\eta_{1R}+\frac{\epsilon_{11}}{4}\right) +i\;\mbox{sin}(P_j)\mbox{sinh}\left(\eta_{1R}+\frac{\epsilon_{11}}{4}\right) \right]e^{i\eta_{1I}}}{4 \mbox{cosh}^2\left(\eta_{1R}+\frac{\epsilon_{11}}{4}\right)+L}, \quad j=1,2,3,~~~~
\label{os}
\eea
where $A_j=e^{{\frac{l_j+\delta_{11}^{(j)}-\epsilon_{11}}{2}}}$, $P_j=\frac{\delta_{11I}^{(j)}-l_{jI}}{2}$, $e^{l_{j}}={\alpha_1^{(j)}}$, $j=1, 2,3$, $L=e^{({R_1-\frac{\epsilon_{11}}{2}})}-2$, $\eta_{1R}=k_{1R}(X-2k_{1I}T)$, $\eta_{1I}=k_{1I}{X}+(k_{1R}^2-k_{1I}^2){T}$, and the other quantities ($\delta_{11}^{(j)}$, $R_1$, and $\epsilon_{11}$) are defined in Eqs.  (\ref{s1s}). Here the amplitude (peak value of the soliton profile) of the polar soliton in $q_j$-th component is 2$A_j,~j=1,2,3$. These PSs have novel types of density profiles like double-hump and flat-top which display interesting features in the soliton interactions, in addition to the standard ``sech" type profile. Here, we have four arbitrary complex parameters for a PS. The density plots of flat-top PS in $q_1$ and $q_3$ components and double-hump PS in $q_2$ component are shown in Fig.~\ref{genpolar} for the parametric choice $k_1 = 1.5-0.3i, \alpha_1^{(1)} =1, \alpha_1^{(2)} = 0.725, \mbox{and} ~\alpha_1^{(3)} = 1$. This double-hump profile has already been reported in Ref. \cite{Ieda} in the context of spinor BECs and also in the context of nonlinear optics \cite{tkjpa}.

\subsection{Bright matter wave two-soliton solution}
\indent The bright two-soliton solution of the integrable three-component Gross-Pitaevskii equations (\ref{ccnls}), obtained by using Hirota's bilinearization method with an auxiliary function, can be written as
\bes\bea
q_j&=&\frac{G^{(j)}}{F},\;\; j = 1,2,3.
\eea
The functions $G^{(j)}$ and $F$ are given by the expressions
\bea
\hspace{-2.5cm}G^{(j)}=&&\alpha _1^{(j)} e^{\eta _1} +\alpha _2^{(j)} e^{\eta _2}+\sum_{u,v=1}^{2} e^{2 \eta _u+\eta _v^{*}+\delta_{uv}^{(j)}}+\sum_{u=1}^{2} e^{\eta_1+\eta_2+\eta_u^*+\delta_u^{(j)}}+\sum_{u,v=1}^{2} e^{2\eta_u+2\eta_v^*+\eta_{3-u}+\mu_{uv}^{(j)}} \nonumber\\
\hspace{-2.5cm}&&+e^{\eta_1+\eta_1^*+\eta_2+\eta_2^*}\left(\sum_{u=1}^{2} e^{\eta_u+\mu_u^{(j)}}+\sum_{u=1}^{2} e^{\eta_1+\eta_2+\eta_u^*+\phi_u^{(j)}}\right),\quad j=1,2,3,~~~\eea\bea
\hspace{-2.5cm}F~=&&1+\sum_{u=1}^{2} e^{\eta_u+\eta_u^*+R_u}+e^{\eta_1+\eta_2^*+\delta_0}+e^{\eta_2+\eta_1^*+\delta_0^*}+\sum_{u,v=1}^{2} e^{2\eta_u+2\eta_v^*+\epsilon_{uv}}+e^{\eta_1^*+\eta_2^*}\sum_{u=1}^{2} e^{2\eta_u+\tau_u}\nonumber\\
\hspace{-2.5cm}&& +e^{\eta_1+\eta_2}\sum_{u=1}^{2} e^{2\eta_u^*+\tau_u^*}+e^{\eta_1+\eta_1^*+\eta_2+\eta_2^*}\left(e^{R_3}+\sum_{u,v=1}^{2} e^{\eta_u+\eta_v^*+\theta_{uv}}+e^{\eta_1+\eta_1^*+\eta_2+\eta_2^*+R_4}\right),~~~~~~
\eea
and the auxiliary function `$s$' is given by
\bea
\hspace{-2.5cm}s=&&\sum_{u=1}^{2} \Gamma_u e^{2 \eta _u}+\Gamma_3 e^{\eta _1+\eta _2} +\sum_{u,v=1}^{2} e^{\eta _u+2\eta _{3-u}+\eta _v^*+\lambda _{uv}} + e^{2 \eta _1+2 \eta _2} \left(\sum_{u=1}^{2} e^{2\eta _u^*+\lambda _u}+e^{\eta_1^*+\eta _2^*+\lambda _3}\right).~~~~
\eea\label{sol2s}\ees
In the above, $\eta_l=k_l(X+ik_lT),~l=1,2,$ and the above two-soliton solution is characterized by eight complex parameters $\alpha_l^{(j)}$ and $~k_l$, where $l=1,2,$ and $j=1,2,3$. The other quantities appearing in Eq. (\ref{sol2s}) are given in Appendix A.

\section{Non-autonomous bright matter wave soliton solution}\label{nonautosol}
The exact bright matter wave solitons of the integrable 3-GP equations (\ref{ccnls}) given in the previous section, and the similarity transformation (\ref{str}) that transforms the non-autonomous 3-GP equation (2) into a standard integrable 3-GP equation (4), pave way to investigate the dynamics of non-autonomous bright matter wave solitons in the presence of different types of time-varying nonlinearity co-efficients as well as for various forms of external potentials. Here, we restrict our study to a particular type of nonlinearity coefficient $c(t)$ having kink-like form, for illustrative purpose. However, it is a straightforward exercise to extend our study to different types of time-varying function for the nonlinearity coefficient with suitable potential modulations determined from the Riccati equation (\ref{ccnls}d).

\subsection{\underline{Kink-like nonlinearity}}
\begin{figure}[h]
\centering\includegraphics[width=0.4\linewidth]{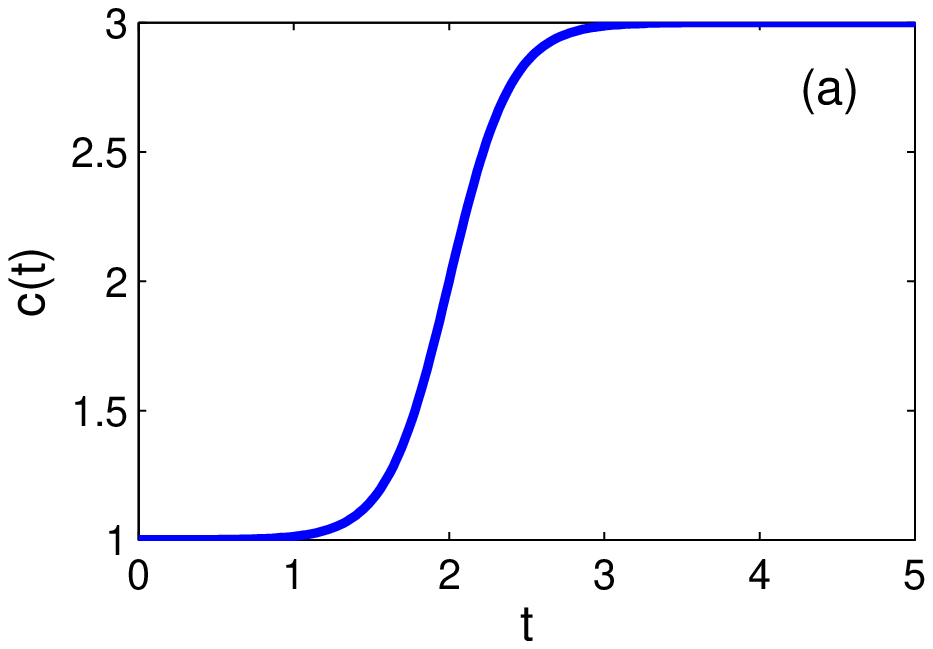}~~\includegraphics[width=0.42\linewidth]{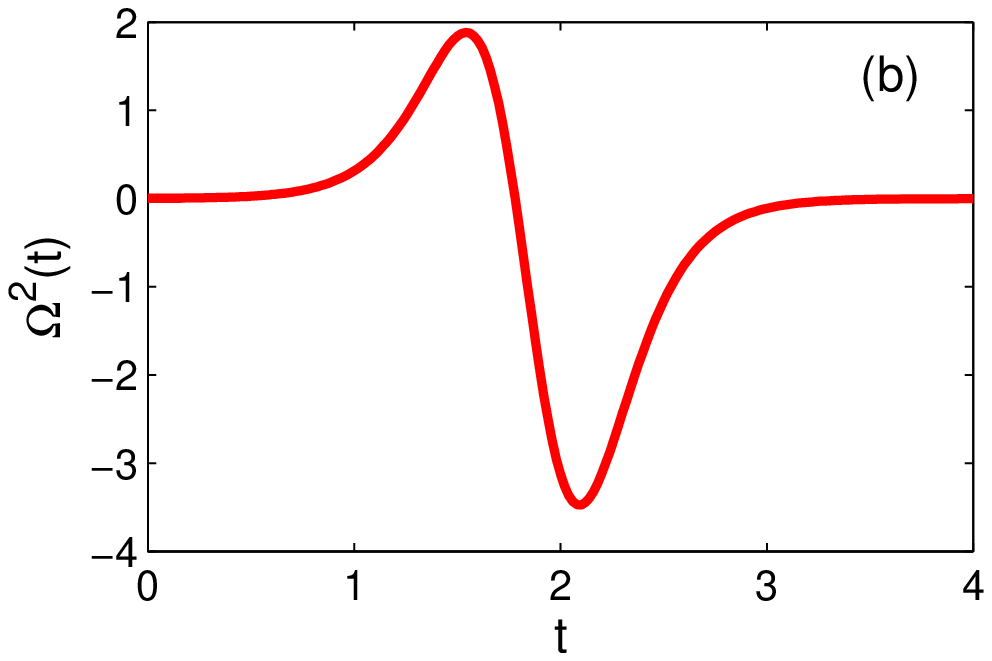}
\caption{(a) Nature of the kink-like nonlinearity and (b) Strength of external harmonic potential, for $\omega=2.5$ and $\delta=-5$.}
\label{potfig}
\end{figure}

\indent Time modulated spin-mixing nonlinearities, particularly fast temporal modulations, display interesting dynamics in spinor BECs \cite{Wen}. A relatively sudden jump in the spin-mixing nonlinearity can be well approximated by the following kink-like form for the nonlinearity coefficient
\bes\bea
c(t) = 2+\mbox{tanh}(\omega t+\delta),
\label{kinklike}
\eea
where $\omega^{-1}$ denotes the time scale characterizing the jump and $\delta$ is an arbitrary constant (see Fig.~\ref{potfig}(a)) with the associated atomic scattering length being $a_{0,2}(t)= \frac{3}{4}a_B[2 + \mbox{tanh}(\omega t + \delta)]$. This form of nonlinearity approximating the temporal modulation by a relatively step-like function can result in a  population growth similar to the experimentally observed condensate growth in BEC \cite{Davis}. Such kink-like nonlinearity has also been considered before in the study of matter wave soliton compression in single component BEC \cite{Abdu} and two-component BECs, but in the absence of spin-mixing effects \cite{similarity,huss,raj}. This type of nonlinearity also finds applications in nonlinear optics, particularly in planar graded-index Kerr-like nonlinear waveguides. The nature of such nonlinearity is shown in Fig.~\ref{potfig}(a). Generally, the sign of time-dependent atomic scattering length can be well tuned from negative to positive with the aid of Feshbach resonance management \cite{micro} and the above choice of time-varying nonlinearity coefficient $c(t)$ (i.e., kink-like nonlinearity) can be a very good candidate for studying the nature of such temporally inhomogeneous spinor BECs. In the present study we have chosen $c(t) $ to range from $1$ to $3$.

\begin{figure}[h]
\centering\includegraphics[width=0.43\linewidth]{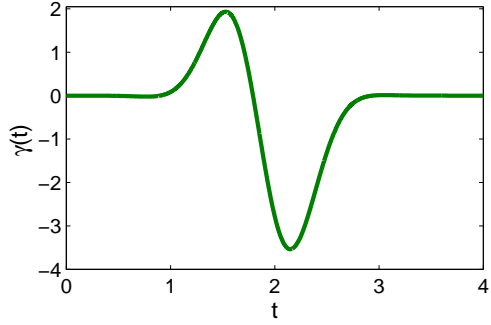}
\caption{Typical plot of $\gamma(t)$  for $w = \sqrt{0.17}$, $C_0 = -1.2$ and $C_3 = 1$.}
\label{hermite}
\end{figure}

The strength of the time-dependent external potential corresponding to the above choice of $c(t)$ (see Eq.~(\ref{kinklike})) is determined from the Riccati equation (\ref{Riccati}) as
\bea
\Omega^2(t) =\frac{-2\omega^2\mbox{sech}(\omega t + \delta)^2[1+2\mbox{tanh}(\omega t + \delta)]}{[2+\mbox{tanh}(\omega t + \delta)]^2}.
\eea\label{pot}\ees
Figure~\ref{potfig}(b) shows that the temporal modulation of the harmonic potential is an asymmetric localized pulse of finite duration, which can be experimentally realized. It is quite interesting to note that the nature of the above function $\Omega^2(t)$ agrees very well with the function
$
\gamma(t) =[C_0H_0(\sigma)+C_3H_3(\sigma)] e^{\frac{-\sigma^2}{2w^2}},
$
where $\sigma = 1.2t-2.25$, $w$ is the width of the Gaussian pulse, $C_0$ and $C_3$ are the coefficients of the zeroth order ($H_0$) and third order ($H_3$) Hermite polynomials, respectively (see Fig.~\ref{hermite}). This function $\gamma(t)$ is nothing but a linear superposition of Hermite-Gaussian (HG) pulse with a Gaussian pulse and can be viewed as a linear superposition of third HG harmonic with the zeroth HG harmonic \cite{Turtsin}.
This type of optical pulse modelling with HG function is a classical concept \cite{ol89} which has obtained renewed attention in the study of dispersion managed solitons \cite{Turtsin}. Indeed, Hermite-Gaussian pulse with suitable modification in its profile can be achieved by means of pulsed lasers and fiber Bragg grating \cite{grat}. We believe that this resemblance of $\Omega^2(t)$ with HG pulse (third harmonic) superimposed by Gaussian pulse (zeroth harmonic), pointed out here, will pave way to realize non-autonomous solitons experimentally not only in spinor condensates but also in multiple species condensates as well as in standard single component condensate. It is also interesting to note that the strength of the potential (see Fig.~\ref{potfig}(b)) and hence the trapping potential $V_{ext}(x,t)$ can admit both negative and positive values. This shows the existence of same type of soliton for both expulsive $(\Omega^2<0)$ and attractive $(\Omega^2>0)$ confining potentials. This is due to the fact that the formation of non-autonomous soliton now depends on time modulation of both the nonlinearity coefficient and the confining potential. This is an advantage of the time modulation of the nonlinearity coefficient as well as trapping potential and has been reported in Ref.\cite{Abdu} for single component BECs and this property can be profitably utilized for creating symbiotic solitons \cite{sym} in spinor condensates.

At this junction, we would like to remark that the converse of the above approach can also be done, i.e., one can fix the time dependence of the strength of the external potential $\Omega^2(t)$, and can determine the corresponding form of $c(t)$  from the Riccati equation. For example, one can consider other physically interesting modulations too for the harmonic potential like, hyperbolic functions \cite{raj}, periodic (Mathieu function) and quasi-periodic potentials \cite{Juan}, optical lattice potential \cite{Juan}, flying-bird potential \cite{Jun}, polynomial function \cite{Del}, etc. Then the corresponding nonlinearity coefficient $c(t)$ can be deduced from Riccati equation (\ref{Riccati}) explicitly, if it is solvable for that particular choice of potential strength. Otherwise, one has to solve the Riccati equation numerically to investigate the dynamics of the non-autonomous solitons.

\subsection{Non-autonomous bright one-soliton solution: Ferromagnetic and Polar solitons}
In this subsection, we explore the influence of the potential $V_{ext}$ (\ref{pot}b) and time-varying kink-like nonlinearity on the matter wave solitons in spinor BECs (\ref{ncgp}). The results are also compared with the standard autonomous spinor BECs of system (\ref{ccnls}) to bring out the salient features of the three component non-autonomous bright matter wave solitons.

The explicit form of the non-autonomous soliton solution for the choice of (\ref{kinklike}) can be constructed from the autonomous soliton solution (\ref{s1s}) with the aid of transformation (\ref{str}). The non-autonomous FS solution can be written as
\bes\bea
q_j&=& \widehat {A}_j~\mbox{sech}\left(\widehat{\eta}_{1R}+R_1/2\right)e^{i(\widehat{\eta}_{1I}+\tilde{\theta})}, \quad j=1,2,3,
\eea
and the non-autonomous PS solution takes the form
\bea
q_j=\frac{2 \widehat{A}_j\left[\mbox{cos}(P_j)\mbox{cosh}\left(\widehat{\eta}_{1R}+\frac{\epsilon_{11}}{4}\right) +i\;\mbox{sin}(P_j)\mbox{sinh}\left(\widehat{\eta}_{1R}+\frac{\epsilon_{11}}{4}\right) \right]e^{i(\widehat{\eta}_{1I}+\tilde{\theta})}}{4 \mbox{cosh}^2\left(\widehat{\eta}_{1R}+\frac{\epsilon_{11}}{4}\right)+L},~j=1,2,3,~
\eea\label{nonatsol}\ees
where, $\widehat{A}_j = A_j\xi_1 \sqrt{2+\mbox{tanh}(\omega t +\delta)}$, $\widehat{\eta}_{1R} = k_{1R}\xi_1[2+\mbox{tanh}(\omega t +\delta)]x - [5t+\frac{1}{\omega}(4\ln[\mbox{cosh}(\omega t +\delta)]-\mbox{tanh}(\omega t +\delta))]2k_{1R}\xi_1^2(\sqrt{2}\xi_1 \xi_2+k_{1I})$, $\widehat{\eta}_{1I} = k_{1I}\xi_1[2+\mbox{tanh}(\omega t +\delta)]x -[5t+\frac{1}{\omega}(4\ln[\mbox{cosh}(\omega t +\delta)]-\mbox{tanh}(\omega t +\delta))]\xi_1^2(2\sqrt{2}\xi_1\xi_2 k_{1I}-k_{1R}^2 + k_{1I}^2)$ and $\tilde{\theta} = \left(\frac{-\omega \mbox{sech}^2(\omega t +\delta)}{2[2+\mbox{tanh}(\omega t +\delta)]}\right)x^2 + 2\xi_1^2\xi_2[2+\mbox{tanh}(\omega t +\delta)]x - 4\xi_2^2\xi_1^4 [5t+\frac{1}{\omega}(4\ln[\mbox{cosh}(\omega t +\delta)]-\mbox{tanh}(\omega t +\delta))]$. The other parameters appearing in (\ref{nonatsol}a) and (\ref{nonatsol}b) are defined below equations (\ref{ms}) and (\ref{os}), respectively. \\

\begin{figure}[h]
\centering\includegraphics[width=0.4\linewidth]{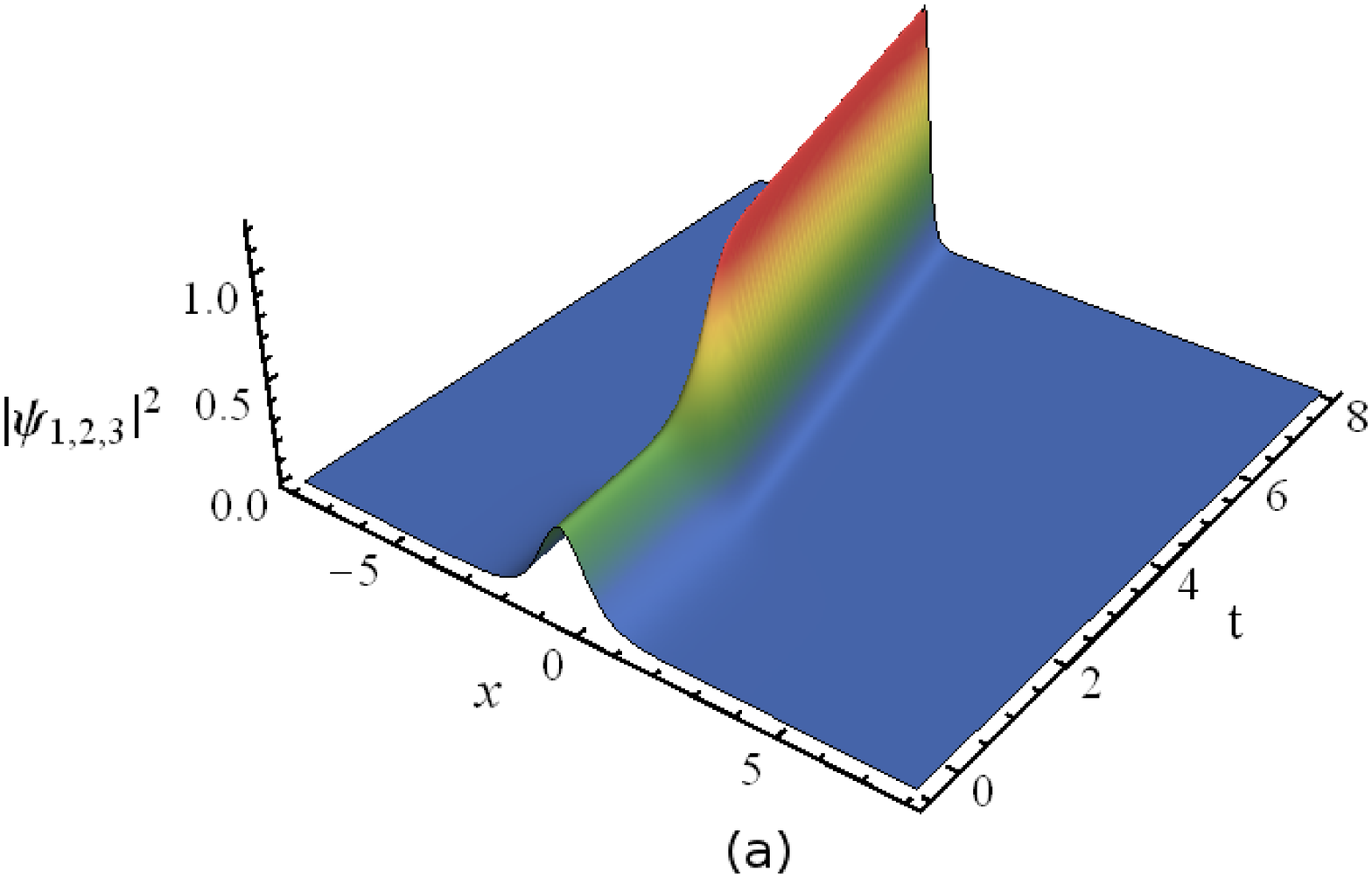}~~~~~~~~~\includegraphics[width=0.35\linewidth]{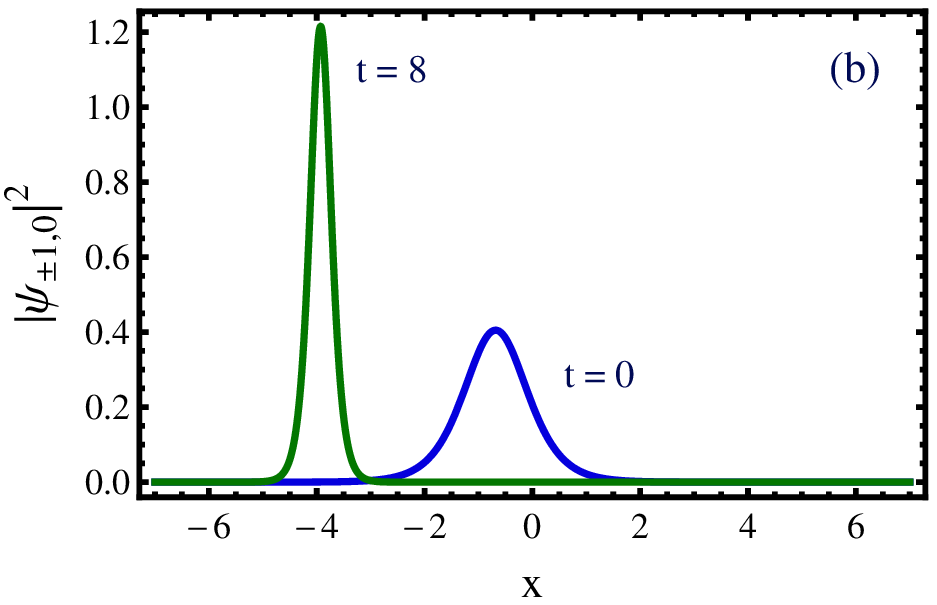}
\caption{(a) Propagation of non-autonomous ferromagnetic soliton (FS) in the 3-GP system (\ref{ncgp}). The parameters are $k_1 = 1.5 - 0.3i$, $\alpha_1^{(1)} = \alpha_1^{(2)} = \alpha_1^{(3)} = 0.2$, $\xi_1 = 0.6$, $\xi_2 = 0.5$, $\delta = -5$ and $\omega = 2.7$. (b) Compression of non-autonomous FS with amplification.}
\label{inhofs}
\end{figure}

\begin{figure}[h]
\centering\includegraphics[width=0.40\linewidth]{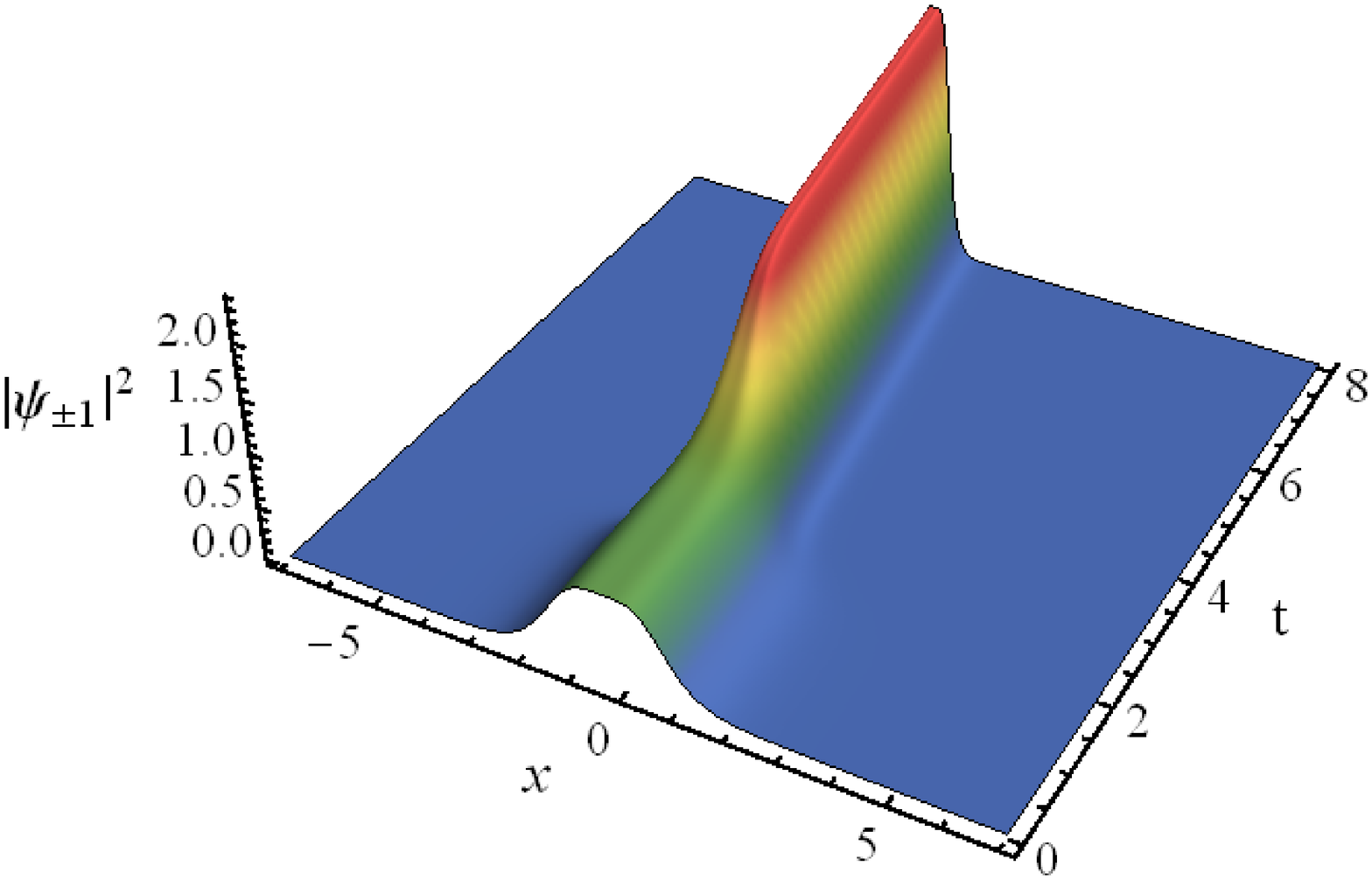}~~~~~~~~~\includegraphics[width=0.4\linewidth]{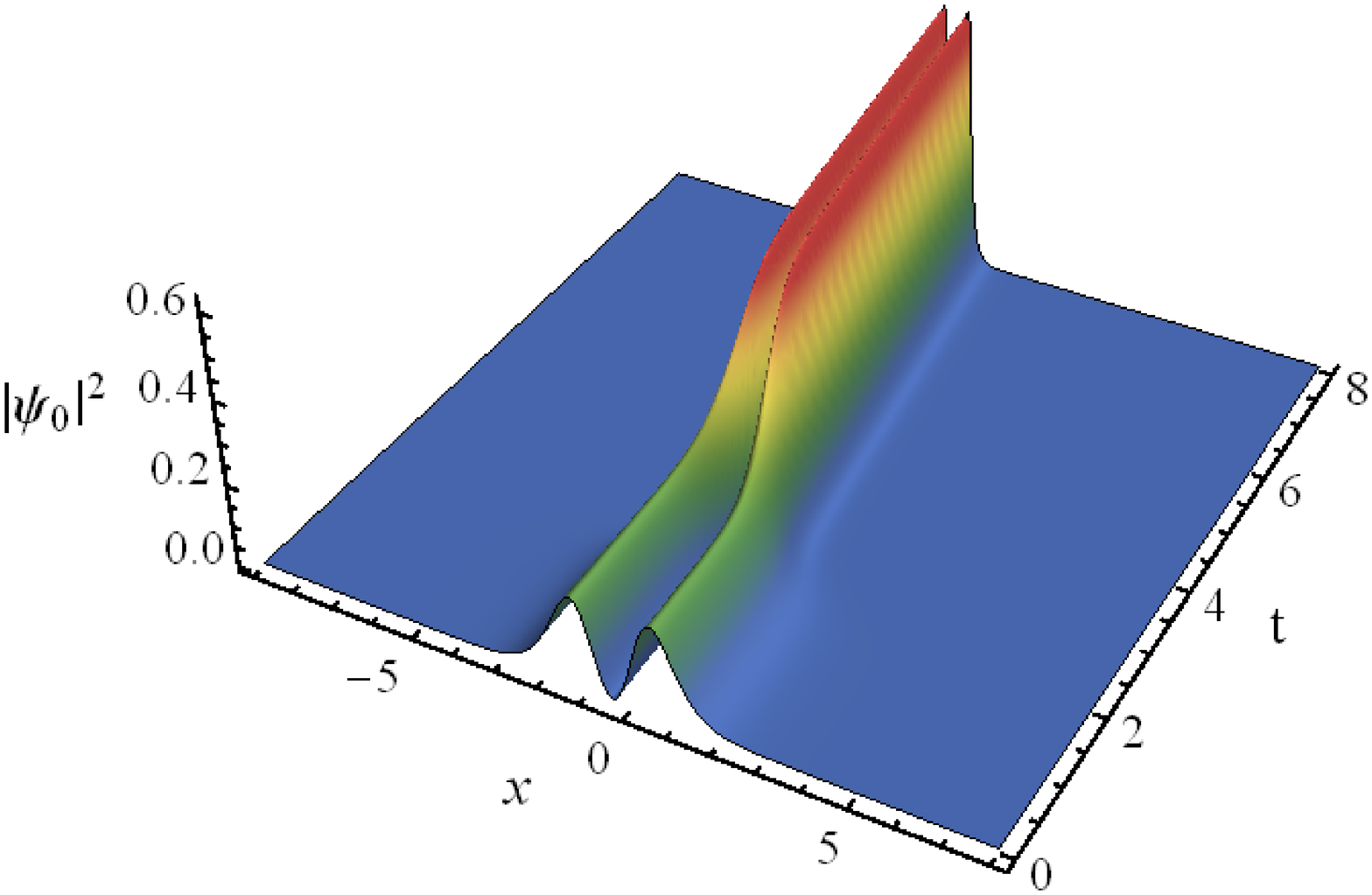}
\caption{Propagation of flat-top (left panel) and double-hump (right panel) non-autonomous polar soliton with kink-like nonlinearity. The parameters are $k_1 = 1.5-0.3i, \alpha_1^{(1)} =1,~ \alpha_1^{(2)} = 0.725,~ \alpha_1^{(3)} = 1$, $\xi_1 = 0.6$, $\xi_2 = 0.5$, $\delta = -5$, and $\omega = 2$.}
\label{inhops}
\end{figure}

The explicit expressions (\ref{nonatsol}a) and (\ref{nonatsol}b) show that the temporal inhomogeneity affects the amplitude, central position and phase of the ferromagnetic and polar solitons in a same manner. This in turn strongly influences the velocity and width of the solitons during propagation. The arbitrary constant $\xi_2$ specifically modulates the central position and also the phase of the soliton. The arbitrary parameter $\xi_1$ modulates the amplitude, as well as central position and phase. As a whole, the kink-like nonlinear function $c(t)$ modulates the soliton profile and introduces a significant step amplification in the soliton density. Also, it affects the width of the solitons to a greater extent. To elucidate the understanding of this particular type of nonlinearity coefficient (\ref{pot}a) and the corresponding potential (\ref{pot}b) we present the non-autonomous FS and PS in Fig.~\ref{inhofs} and Fig.~\ref{inhops}, respectively.

The propagation of non-autonomous FS and its two dimensional plot (at $t=0$ and $t =8$) are depicted in Fig.~\ref{inhofs}(a) and Fig.~\ref{inhofs}(b), respectively. We observe that the amplitude and hence the density of FS are increased significantly. More importantly, the width of the FS is narrowed down, as inferred from the $2D$ plot Fig.~\ref{inhofs}(b). At $t=0$, the soliton is wider but with smaller amplitude, whereas at $t =8$ the width of the soliton gets compressed and its amplitude is increased. This kind of pulse compression is one of the challenging tasks in information transfer using solitons. This property suggests the bright non-autonomous matter wave solitons in spinor condensates to be a suitable candidate for information transfer and also for quantum information process. The central position of the soliton is also shifted significantly due to the $\xi_2$ parameter.

Similarly, the propagation of non-autonomous PS is shown in Fig.~\ref{inhops}, where the flat-top (double-hump) nature of the soliton profile is unaffected due to temporal nonlinearity coefficient. The width of the soliton profiles are much compressed, but still they retain their double-hump/flat-top shapes. Here also the compression of soliton resulting in soliton amplification and shift in the central position takes place.

\subsection{Non-autonomous bright two-soliton solution}
The explicit form of non-autonomous bright two-soliton solution of the system (\ref{ncgp}) can be written as
\bea
\hspace{-1.0cm} &&(\psi_{+1},~\psi_{0},~\psi_{-1})^T = \xi_1\sqrt{2+\mbox{tanh}(\omega t +\delta)}~e^{i\tilde{\theta}(x,t)} \left(\frac{G^{(1)}}{F},~\frac{G^{(2)}}{F},~\frac{G^{(3)}}{F}\right)^T, \label{nonat2sol}
\eea
where $\frac{G^{(j)}}{F},~j=1,2,3,$ is same to that of autonomous case (\ref{sol2s}b)--(\ref{sol2s}d) with the redefinition of $\eta_l = \widehat{\eta}_l$. Here $\widehat{\eta}_l = k_{l}\xi_1[2+\mbox{tanh}(\omega t +\delta)]x - [5t+\frac{1}{\omega}(4\ln[\mbox{cosh}(\omega t +\delta)]-\mbox{tanh}(\omega t +\delta))]\xi_1^2k_l(2\sqrt{2}\xi_1\xi_2- i k_l),~ l =1,2,$ and
$\tilde{\theta} = \left(\frac{-\omega \mbox{sech}^2(\omega t +\delta)}{2[2+\mbox{tanh}(\omega t +\delta)]}\right)x^2 + 2\xi_1^2\xi_2[2+\mbox{tanh}(\omega t +\delta)]x - 4\xi_2^2\xi_1^4 [5t+\frac{1}{\omega}(4\ln[\mbox{cosh}(\omega t +\delta)]-\mbox{tanh}(\omega t +\delta))]$.

\section{Interaction of non-autonomous bright matter wave solitons}\label{collisions}
In this section, we investigate different types of soliton interaction scenario in the presence of kink-like nonlinearity and compare with the autonomous soliton interactions in system (\ref{ccnls}). The interaction scenario of bright solitons in autonomous 3-GP equations (\ref{ccnls}) have been classified into three categories in Ref. \cite{Ieda}, namely (i) collision between two FSs, (ii) collision of FS with PS and (iii) collision among two PSs. There itself, it has been shown that the collision between PS is a purely elastic collision (like collision of solitons in scalar NLS system) and during its collision with a FS, the FS induces switching in the PS solitons but FS remains intact. The third collision process among FSs is further interesting. It has been pointed out that during the collision among FS there occurs spin precession (spin of each soliton moves on a circumference around the total spin axis, i.e., spin of each soliton gets rotated) resulting as a consequence of total spin conservation \cite{Ieda}. Additionally, in the present work, we have identified that for a particular choice of the parameter $\alpha_i^{(j)}$, $i=1,2,~j=1,2,3$, associated with the spin polarization, the FSs can also exhibit elastic collision scenario too as that of polar solitons. Based on this, we include this elastic collision scenario of FSs in addition to the three types of collisions identified in Ref. \cite{Ieda} and summarize them in Table 1. The $\Gamma_j$'s, $j=1,2,3$, appearing in Table 1 are defined as $\Gamma_1 = \alpha_1^{(1)}\alpha_1^{(3)}-(\alpha_1^{(2)})^2,~\Gamma_2 = \alpha_2^{(1)}\alpha_2^{(3)}-(\alpha_2^{(2)})^2$ and $\Gamma_3 = \alpha_1^{(1)}\alpha_2^{(3)}+\alpha_2^{(1)}\alpha_1^{(3)}-2\alpha_1^{(2)}\alpha_2^{(2)}$.

\begin{table}[h]
\label{table1}
\caption{\label{tablel}Possible interactions between two bright matter wave solitons based on the choice of $\Gamma_1$, $\Gamma_2$ and $\Gamma_3$.}
\begin{center}
\begin{tabular}{@{}cccccc}
\hline
Case & $\Gamma_1$ & $\Gamma_2$ & $\Gamma_3$ & soliton $S_1$ & soliton $S_2$\\
\hline
(i) & 0 & 0 & 0 & FS:  E & FS:  E\\ 
(ii) & 0 & 0 & $\neq 0$ & FS: SP & FS: SP\\
(iii) & 0 & $\neq 0$ & $\neq 0$ & FS:  E & PS: SW\\
(iv) & $\neq 0$ & $\neq 0$ & $\neq 0$ & PS:  E & PS:  E\\
\hline
\end{tabular}
\end{center}
\end{table}

In Table 1, the abbreviations `E', `SP' and `SW' represent the elastic collision, spin precession and spin-switching interactions, respectively. Indeed, the first two cases (i.e., case (i) and case (ii)) are sub-cases of the broad category corresponding to the interaction of two FSs. In the above, we have identified the type of the two interacting solitons (as either FS or PS) by calculating their total spin explicitly from the asymptotic expressions of those solitons obtained from the exact two soliton solution (\ref{sol2s}). The detailed expressions of the asymptotic analysis corresponding to the above said four cases are given in Appendix B.

\subsection{\textbf{Interaction between non-autonomous FSs} }
As a prelude, we discuss the interaction between two FSs described by the two-soliton solution (\ref{sol2s}) of autonomous system (\ref{ccnls}), resulting for the choices $\Gamma_1=0$ and $\Gamma_2=0$. The amplitude of FSs $(A_j$) after interaction can  be related to that of before interaction using the asymptotic expressions given in Appendix B, as
\bes\bea
A_j^{1+}=T_j^{(1)}~A_j^{1-}, \qquad A_j^{2+}=T_j^{(2)}~A_j^{2-}, \quad j=1,2,3,
\eea
where the superscript (subscript) denotes the soliton (spin component) number and the $-$ ($+$) sign appearing in the superscript denotes the soliton before (after) interaction. Here, the transition amplitudes $T_j^{(1)}$ and $T_j^{(2)}$ are given by
\bea
T_j^{(1)}&=& \frac{\left(\chi_1 +\chi_3-1 \right)}{\sqrt{1-\chi_1 \chi_2+\chi_5}}\left(\frac{(k_1^*+k_2)(k_1-k_2)}{(k_1+k_2^*)(k_1^*-k_2^*)}\right)^{\frac{1}{2}},\\
T_j^{(2)}&=& \frac{\left(1-\chi_2 +\chi_4 \right)}{\sqrt{1-\chi_1 \chi_2+\chi_5}}\left(\frac{(k_1^*+k_2)(k_1-k_2)}{(k_1+k_2^*)(k_1^*-k_2^*)}\right)^{\frac{1}{2}}, ~~ j=1,2,3,
\eea  \label{intfsa}\ees
where $\chi_1=\frac{\alpha_2^{(j)} \kappa_{12}}{\alpha_1^{(j)} \kappa_{22}}$, $\chi_2=\frac{\alpha_1^{(j)} \kappa_{21}}{\alpha_2^{(j)} \kappa_{11}} $, $\chi_3=\frac{ \Gamma_3 \alpha_2^{(4-j)*}}{(k_1-k_2)\alpha_1^{(j)} \kappa_{22}}$, $\chi_4= \frac{ \Gamma_3 \alpha_1^{(4-j)*}}{(k_1-k_2)\alpha_2^{(j)} \kappa_{11}}$, $j=1,2,3$ and $\chi_5=\frac{ |\Gamma_3|^2}{|k_1-k_2|^2 \kappa_{11} \kappa_{22}}$. As mentioned earlier, the above asymptotic expressions can be used for identifying two types of soliton interactions, namely the interaction with and without spin precession (i.e., case (i) and case (ii) in Table 1). Indeed, the transition amplitudes defined here and in the following determine whether there occurs switching of condensates or not. The transition amplitudes also act as measure of switching efficiency.\\

\begin{figure}[h]
\centering\includegraphics[width=0.45\linewidth]{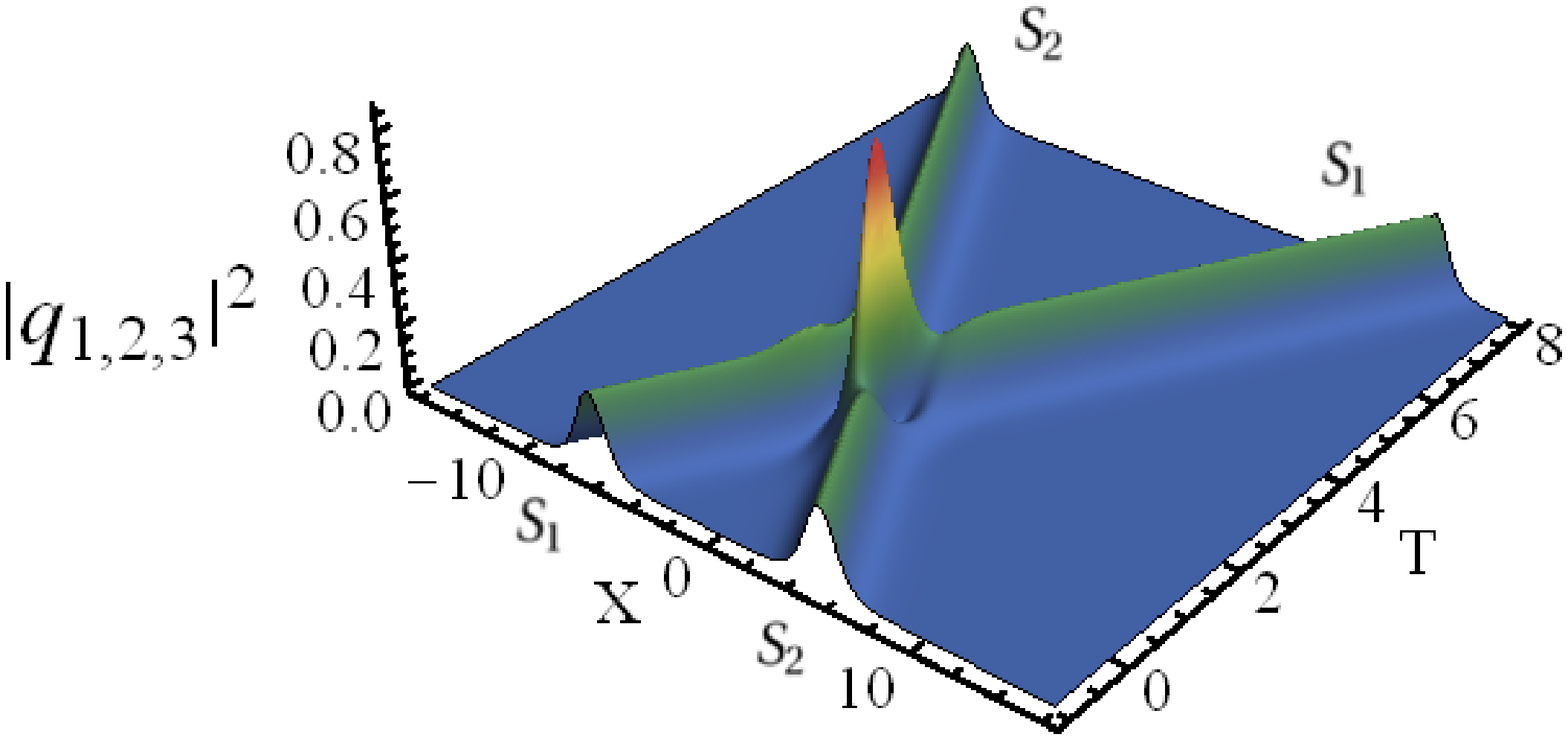}~~~~~~\includegraphics[width=0.45\linewidth]{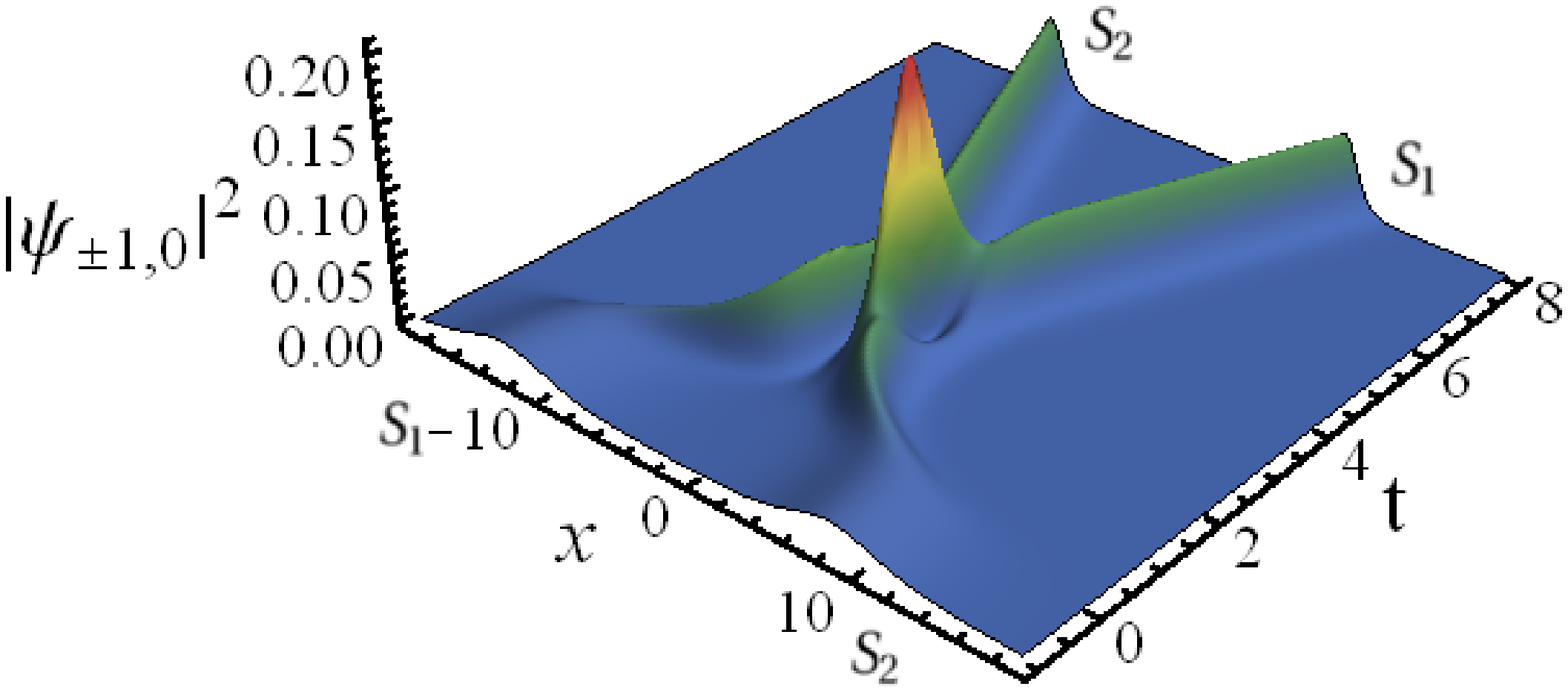}
\caption{Elastic interaction of two autonomous FSs given by (\ref{sol2s}) and non-autonomous FSs given by (\ref{nonat2sol}). The soliton parameters are $k_1 = -1 + i$, $k_2 = 1 - i$, $\alpha_1^{(1)} = \alpha_1^{(2)} = \alpha_1^{(3)} =0.03$, $\alpha_2^{(1)} = \alpha_2^{(2)} = \alpha_2^{(3)} = 0.04$, $\xi_1 = 0.2$, $\xi_2 = 0$, $\delta = -1$, and $\omega = 2.5$.}
\label{intfsfig1}
\end{figure}

\noindent \textbf{Case (i):} When $\Gamma_1=\Gamma_2=\Gamma_3=0$, the transition amplitudes become unimodular, revealing the fact that the ferromagnetic solitons exhibit elastic collision thereby retaining their spin polarization, amplitude and velocity after interaction. There occurs only a phase-shift after interaction which can be defined as $\Phi_1=\frac{R_3-R_2-R_1}{2k_{1R}} \equiv \frac{1}{2k_{1R}}\ln \left[\frac{|k_1-k_2|^2}{|k_1+k_2^*|^2} \left(1-\chi_1 \chi_2\right)\right]$ for FS($S_1$) and $\Phi_2=\frac{k_{1R}}{k_{2R}}\Phi_1$ for FS($S_2$). In the left panel of Fig.~\ref{intfsfig1}, we present such an elastic type of interaction of FSs in 3-GP system (\ref{ccnls}) and the corresponding parameters are given in the figure caption. The figure clearly shows that there is no spin rotation for the colliding solitons in all the three spin components. This means that the spin precession (spin rotation of individual solitons) noticed in Ref.\cite{Ieda} does not occur in the FSs for this choice. This type of elastic interaction of two autonomous bright FSs has not been pointed out in earlier works on spinor condensates\cite{Ieda} to the best of our knowledge.

The asymptotic analysis of the non-autonomous two-soliton solution (\ref{nonat2sol}) shows that the transition amplitudes for non-autonomous FSs will take the same form as that of Eq.(\ref{intfsa}). Hence the nature of soliton interaction is unaffected by the inclusion of time-varying nonlinearity and potential which satisfy (\ref{pot}b). This is shown in the right panel of Fig.~\ref{intfsfig1}. We observe that the density profiles of the two solitons are modulated by the kink-like nonlinearity. As a result of this, there is a uniform increase in the amplitude and suppression of width in both the solitons after collision. However, there is no spin precession of solitons in the three components. The kink-like time-varying nonlinearity coefficient results in noticeable reduction in the separation distance between the solitons before and after collision. This is contrary to the collision of bright solitons in 3-GP system (\ref{ccnls}), depicted in left panel of Fig.~\ref{intfsfig1}, where the relative separation distance increases after collision. The relative separation distance between the two non-autonomous FSs ($S_1$ and $S_2$) before and after interaction can be written as $t_{12}^-=\frac{k_{1R}R_2-k_{2R}R_1}{2k_{1R}k_{2R}} $ and $t_{12}^+=\frac{k_{1R}(R_3-R_1)-k_{2R}(R_3-R_2)}{2k_{1R}k_{2R}} $, respectively. Hence the change in relative separation distance becomes $\Delta t_{12}=t_{12}^- - t_{12}^+ \equiv \left(1-\frac{k_{1R}}{k_{2R}} \right)\Phi_1$. This relative separation distance plays crucial role in the context of soliton complexes \cite{akhm}. Another main difference is that the autonomous FSs collide sooner than the non-autonomous FSs.\\

\begin{figure}[h]
\centering\includegraphics[width=0.33\linewidth]{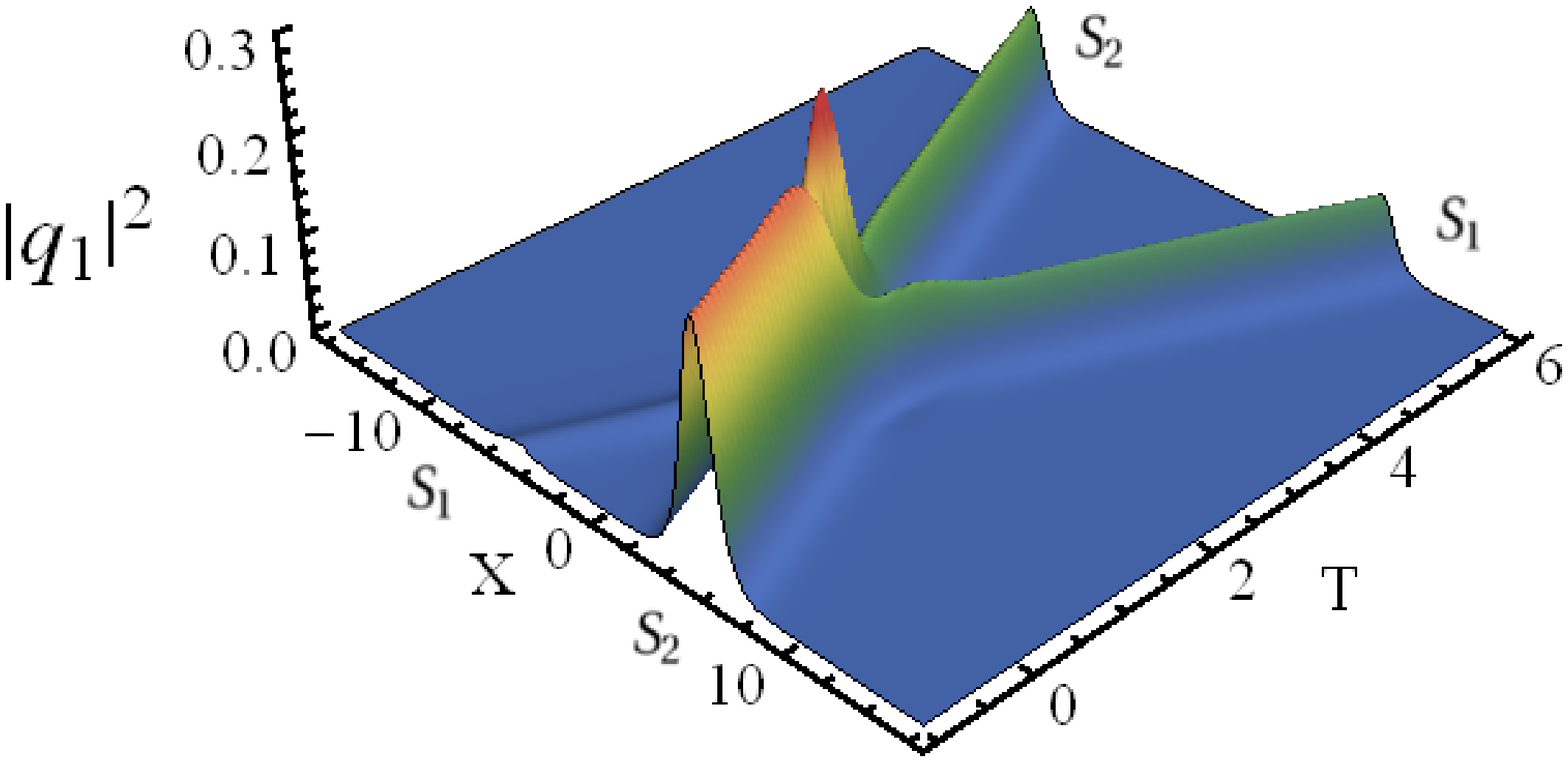}\includegraphics[width=0.33\linewidth]{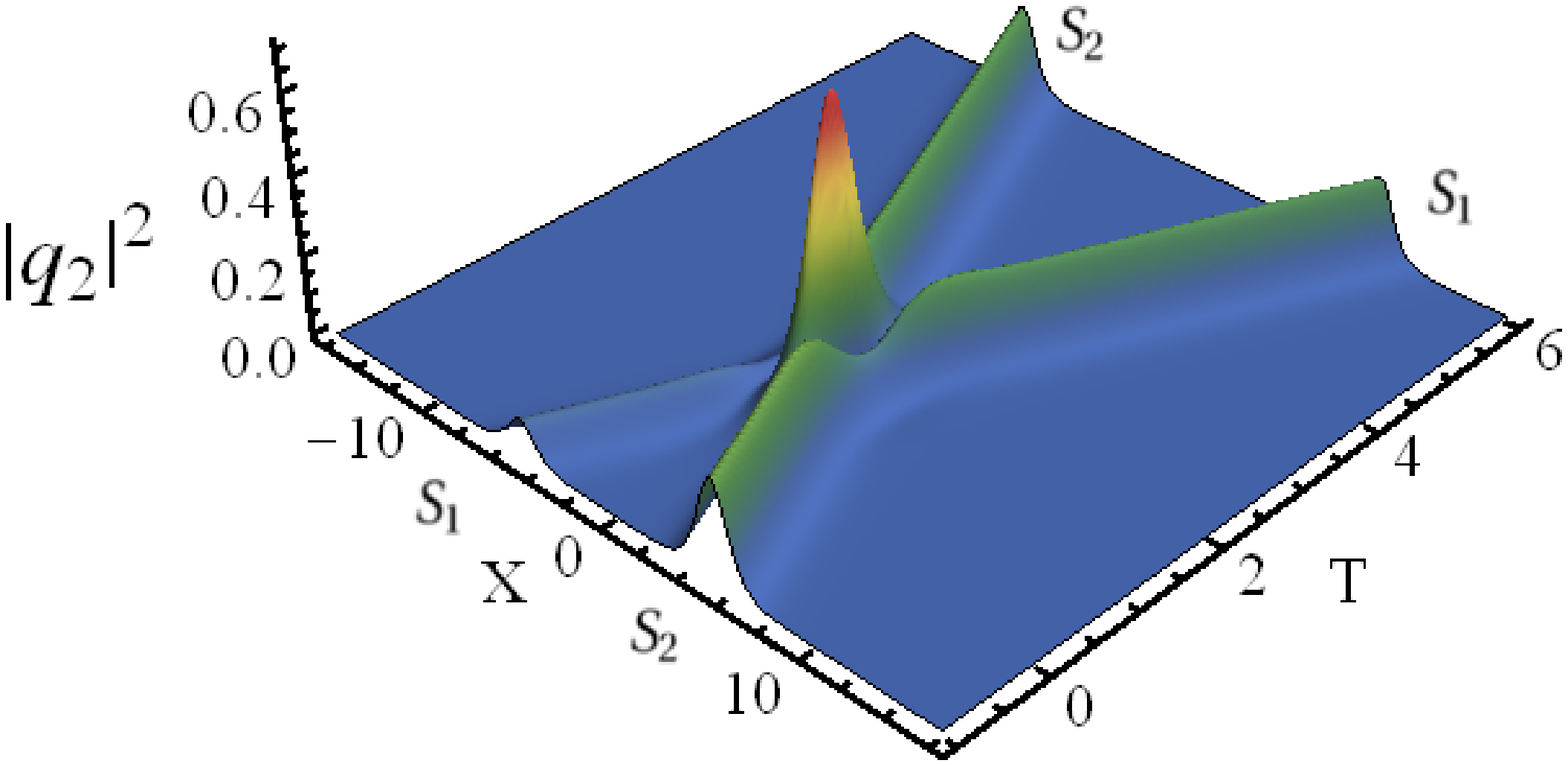}\includegraphics[width=0.33\linewidth]{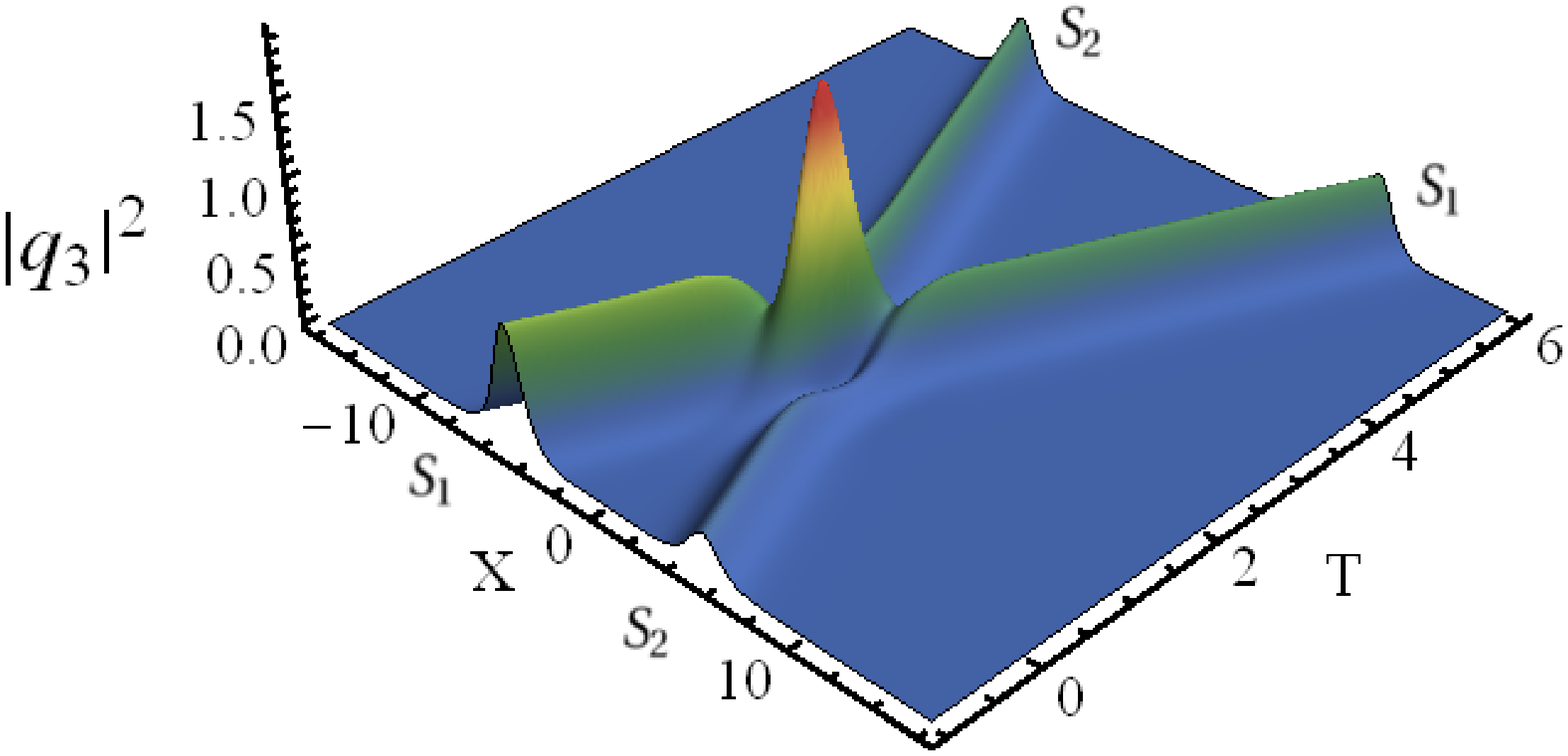}\\
\includegraphics[width=0.33\linewidth]{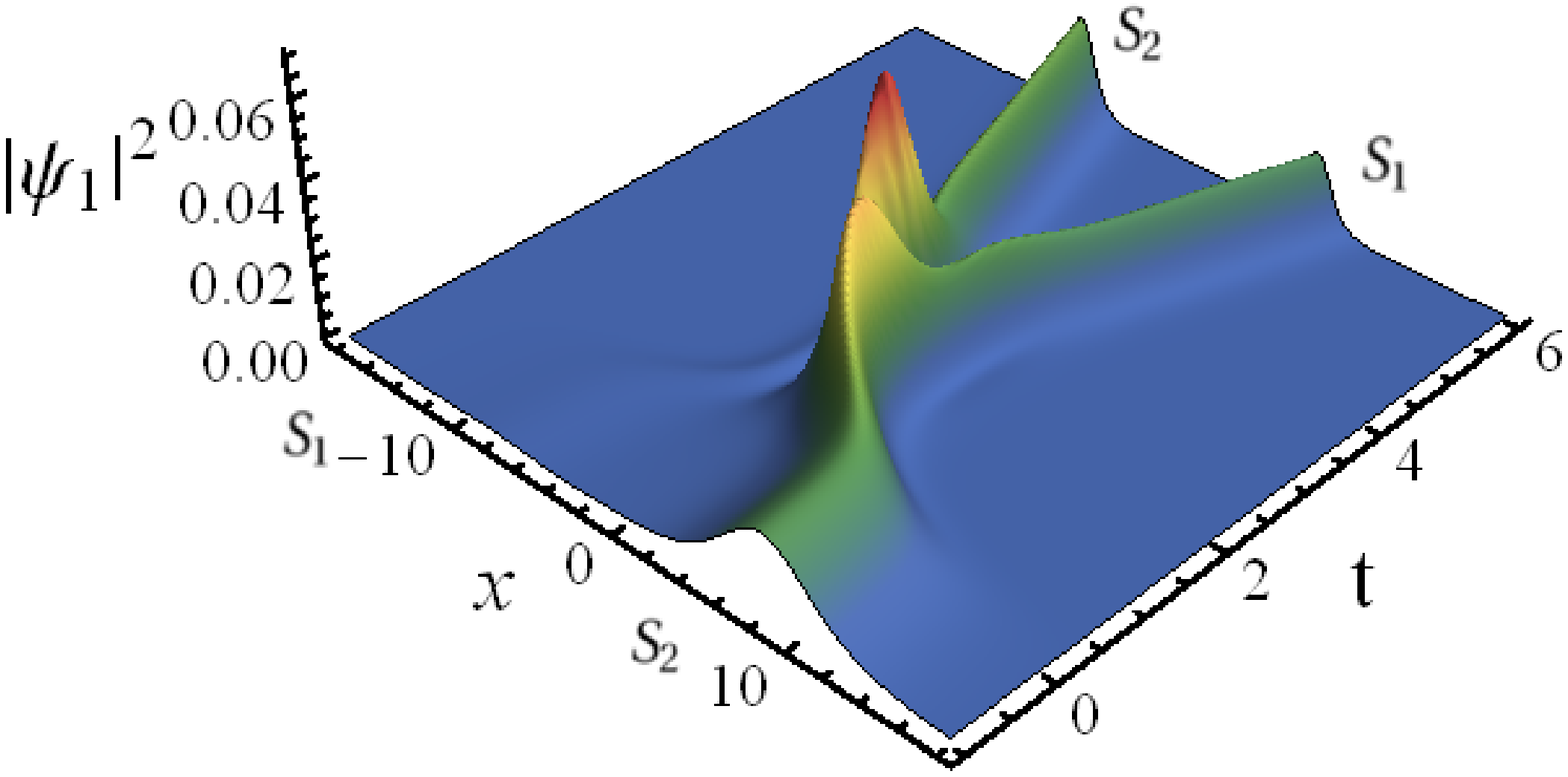}\includegraphics[width=0.33\linewidth]{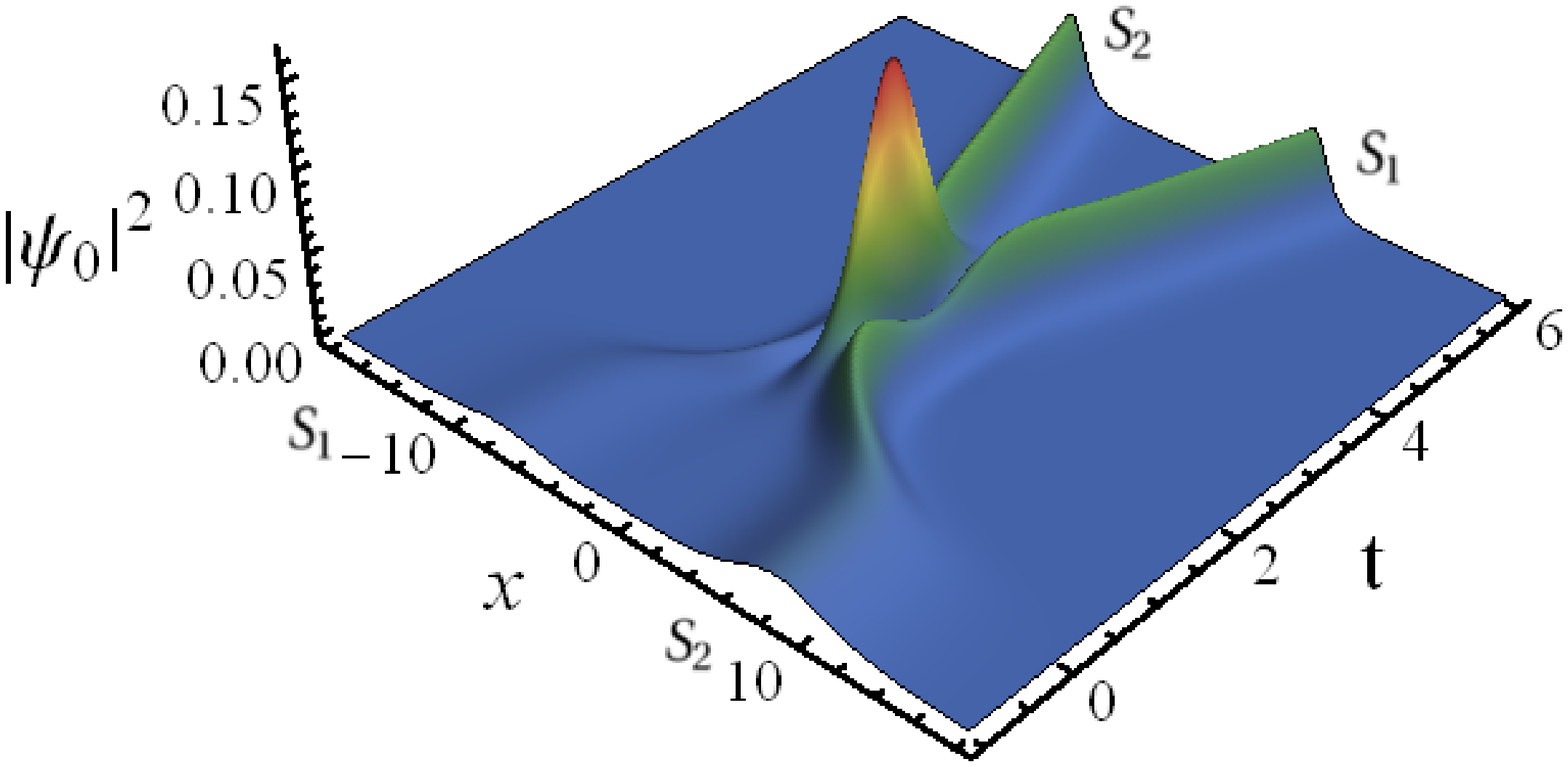}\includegraphics[width=0.33\linewidth]{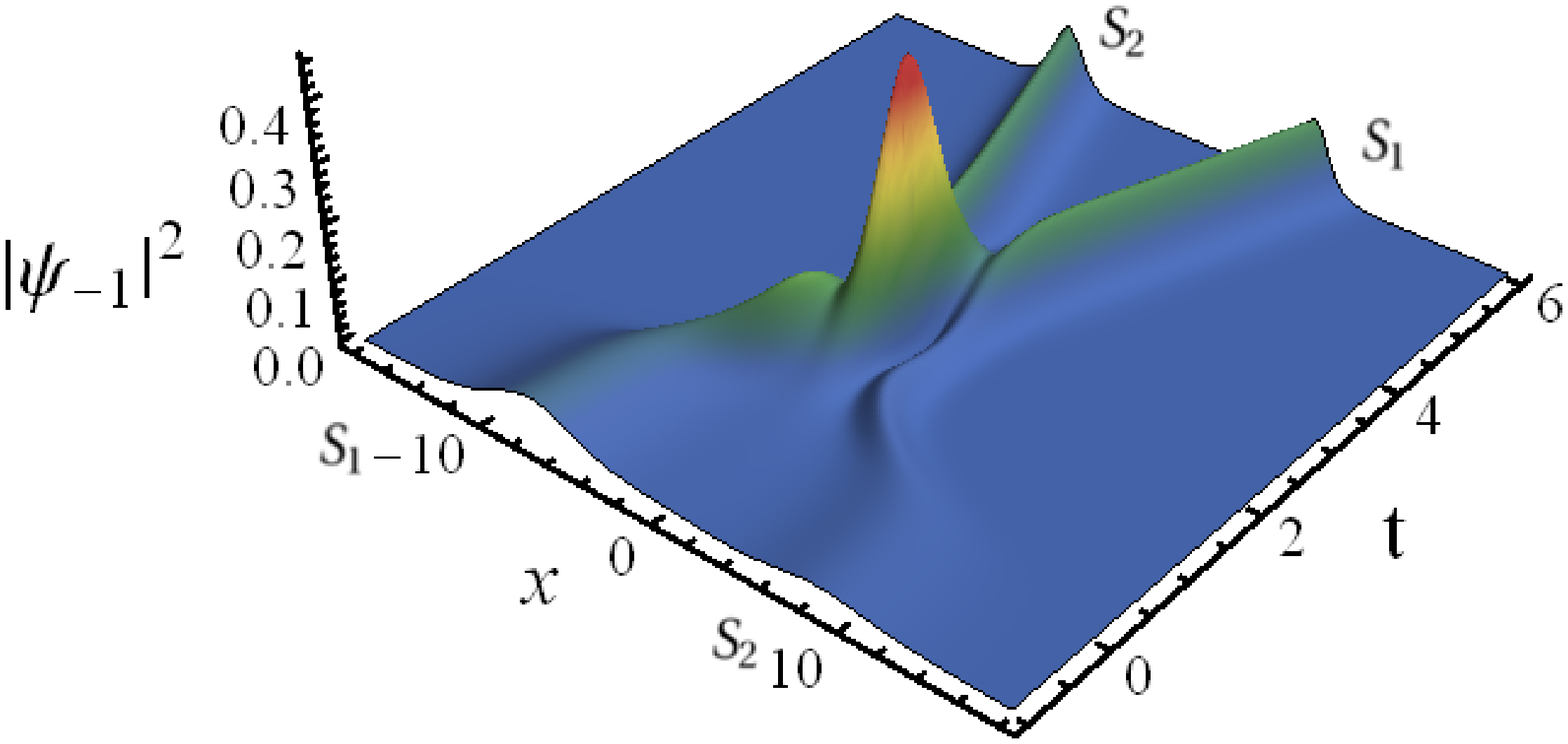}
\caption{Shape-changing interaction of two FSs in autonomous 3-GP system (\ref{ccnls}) (top panels) and in non-autonomous 3-GP system (\ref{ncgp}) with time-varying kink-like nonlinearity coefficient and inhomogeneous potential given by (\ref{pot}) (bottom panels). The soliton parameters are $k_1 = -1+ i$, $k_2 = 1 - i$, $\alpha_1^{(1)} =0.01,~\alpha_1^{(2)} = 0.03,~ \alpha_1^{(3)} = 0.09$, $\alpha_2^{(1)} = \alpha_2^{(2)} = \alpha_2^{(3)} = 0.04$, $\xi_1 = 0.5$, $\xi_2 = 0$, $\delta = -2$, and $\omega  = 2.7$.}
\label{intfsfig2}
\end{figure}

\noindent \textbf{Case (ii)} $\Gamma_1=\Gamma_2=0,~\Gamma_3 \neq 0$: For this case, it can be verified from Eq.~(\ref{intfsa}) that the transition amplitudes can never become unimodular. Hence the amplitudes of the two FSs do alter after interaction. This leads to the spin precession resulting in the suppression and enhancement of density of both solitons due to collision accompanied by phase-shift $\Phi_1=\frac{R_3-R_2-R_1}{2k_{1R}} \equiv \frac{1}{2k_{1R}}\ln \left[\frac{\Gamma_3 \Gamma_3^*+|k_1-k_2|^2}{|k_1+k_2^*|^2} \left(1-\chi_1 \chi_2\right)\right]$ for FS($S_1$) and $\Phi_2=\frac{k_{1R}}{k_{2R}}\Phi_1$ for FS($S_2$), as demonstrated in Ref.\cite{Ieda}. Figure~\ref{intfsfig2} shows a typical collision of two FSs with spin precession for the autonomous (top panels) and the non-autonomous (bottom panels) solitons. Here also we observe similarities and differences between the autonomous FSs collision and collision of non-autonomous FSs.

The top panels of Fig.~\ref{intfsfig2}, displays collision between bright solitons in the autonomous 3-component GP equation (\ref{ccnls}) in the ($X$-$T$) plane. The figure shows that, due to spin precession there occurs inter-and intra-component switching of condensates. It should be noticed here that the number density in individual spinor state is non-conserved. However, the total number density is conserved. We would like to remark that, during performing plotting, we have identified for certain parametric choices the number density in every component is almost conserved. Note that, this kind of spin precession interaction between the two FSs is not at all possible in the two-component spinor condensates (i.e., two-component GP equation resulting from (\ref{ccnls}) with $\psi_{+1}=\psi_{-1}$), as the choice $\Gamma_1=\Gamma_2=0$ will always result in $\Gamma_3=0$.

Now it is of interest to raise the question how this collision scenario is affected in the presence of temporally varying nonlinearity coefficient and driving potential that still leaves the non-autonomous GP equation solvable. To answer this question, we plot the non-autonomous spinor bright soliton collision occurring in the presence of kink-like temporal nonlinearity and external potential of the form (\ref{pot}b), in the lower panels of Fig.~\ref{intfsfig2}.

We observe that the temporal nonlinearity merely modulates the amplitude and changes the width of the colliding solitons but the switching nature of spinor solitons (spin precession) remains unaltered. However the two interacting non-autonomous solitons spend much time in the interaction regime. Significant reduction in the relative separation distance between the two colliding non-autonomous ferromagnetic solitons also takes place after collision. To elucidate the understanding of this collision and to compare the amount of condensate switching in a particular component in the systems (\ref{ncgp}) and (\ref{ccnls}), we present the results of the asymptotic analysis of the autonomous and non-autonomous bright two-soliton solution, respectively, in appendices B and C. In fact, the transition amplitude $(T_j^{(l)} = \frac{A_j^{l+}}{A_j^{l-}},~j=1,2,3,~l=1,2)$ gives the measure of switching of condensates due to spin precession. From the asymptotic expressions (given in Appendix C) we find that the transition amplitudes of non-autonomous FS($S_1$) and FS($S_2$) are
$T_j^{(1)}= \frac{\left(\chi_1 +\chi_3-1 \right)}{\sqrt{1-\chi_1 \chi_2+\chi_5}}\left(\frac{(k_1^*+k_2)(k_1-k_2)}{(k_1+k_2^*)(k_1^*-k_2^*)}\right)^{\frac{1}{2}}$,
$T_j^{(2)}= \frac{\left(1-\chi_2 +\chi_4 \right)}{\sqrt{1-\chi_1 \chi_2+\chi_5}}\left(\frac{(k_1^*+k_2)(k_1-k_2)}{(k_1+k_2^*)(k_1^*-k_2^*)}\right)^{\frac{1}{2}}$, $j=1,2,3,$ respectively. This is exactly the same as that of autonomous solitons in (\ref{ccnls}) which can be obtained by a standard asymptotic analysis following the lines of references \cite{tkjpa,tkopt,tkpra}. Thus, the switching nature of solitons for non-autonomous solitons of system (\ref{ncgp}), for the kink-like nonlinearity, is exactly same to that of autonomous solitons in system (\ref{ccnls}). Hence the advantage of tuning temporal nonlinearity by Feshbach resonance results in a wider range of coupling coefficient $c(t)$ for which the interaction of FSs with spin precession can be realized, with same switching efficiency of the standard integrable 3-GP system (\ref{ccnls}). Another main advantage of the introduced kink-like nonlinearity lies in its ability to tune the amplitudes, width, position and propagation direction of interacting solitons suitably by altering the parameters $\omega$ and $\delta$.

\subsection{\textbf{Interaction between non-autonomous FS and PS}}
The interaction between non-autonomous FS and PS, arises for the parametric choices $\Gamma_1=0$ and $\Gamma_2 \neq 0$, respectively, as given by case (iii) in Table 1. We notice that during its collision with non-autonomous FS, the non-autonomous PS undergoes spin-switching leaving the FS unaltered as in the integrable autonomous 3-GP system (\ref{ncgp}) \cite{Ieda}. This implies that the temporal dependence of $c(t)$ does not alter the nature of solitons as in the case of interaction of two FSs.
\begin{figure}[h]
\centering\includegraphics[width=0.4\linewidth]{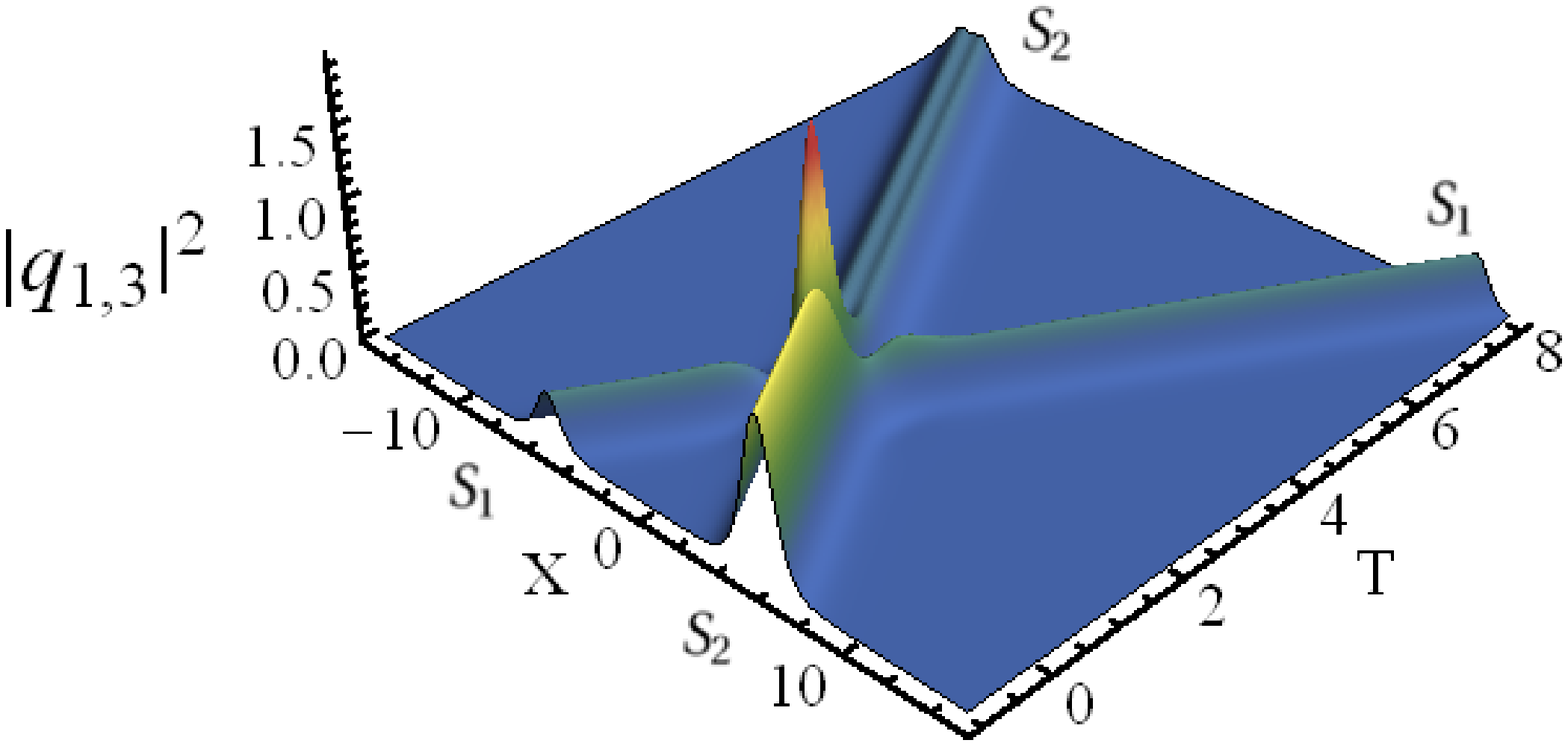}~~~~~~~\includegraphics[width=0.4\linewidth]{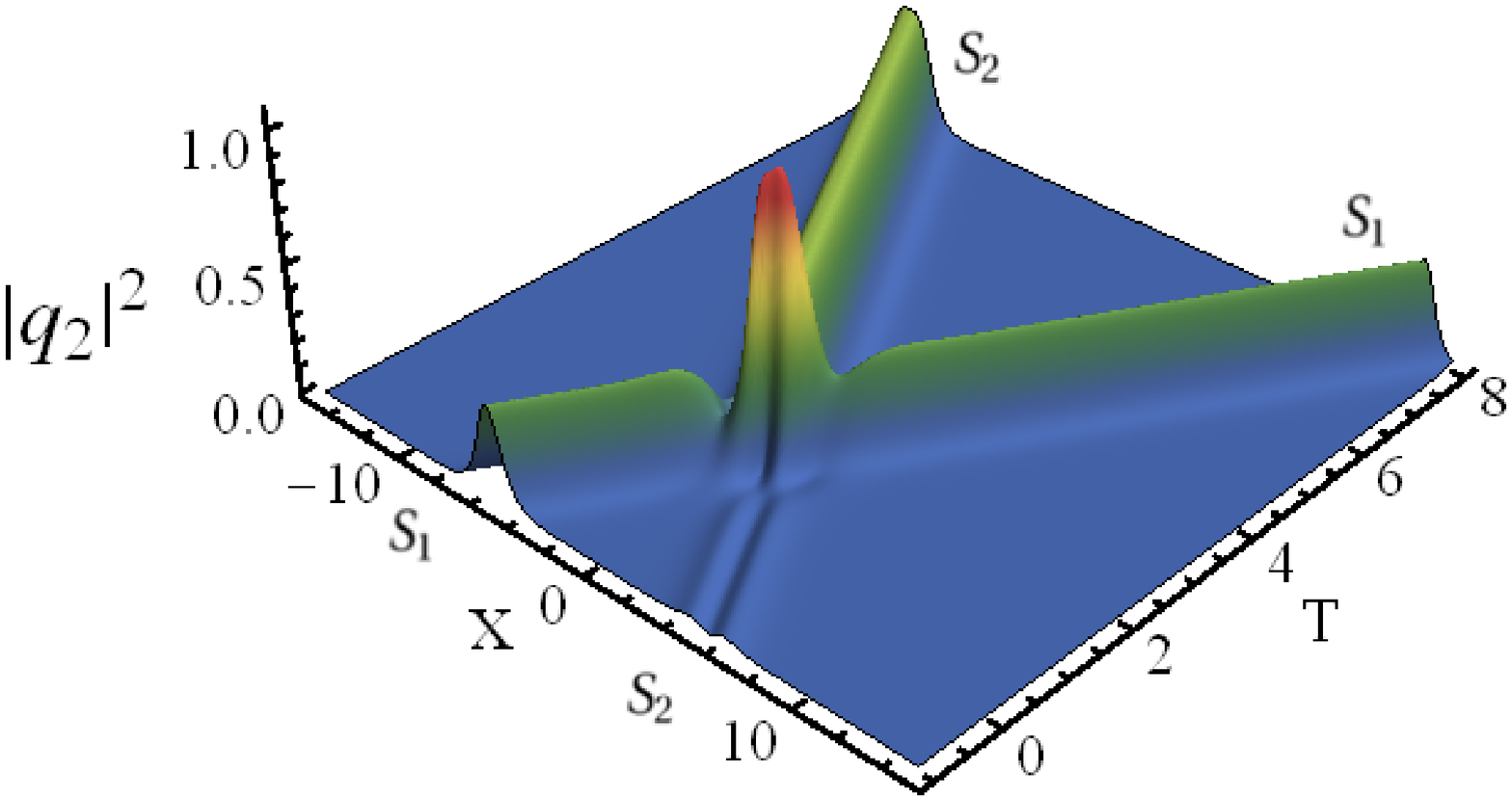}\\
\includegraphics[width=0.4\linewidth]{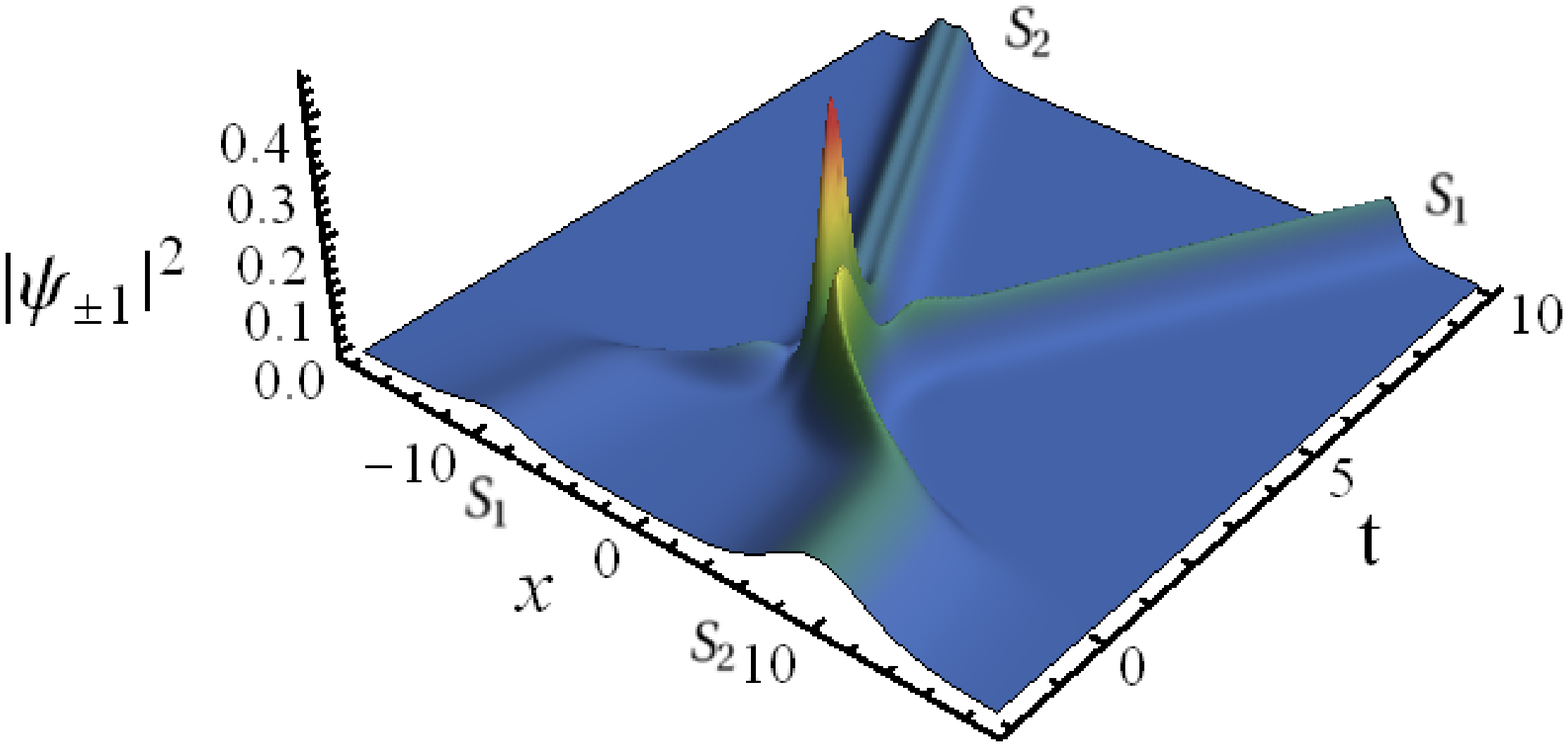}~~~~~~~\includegraphics[width=0.4\linewidth]{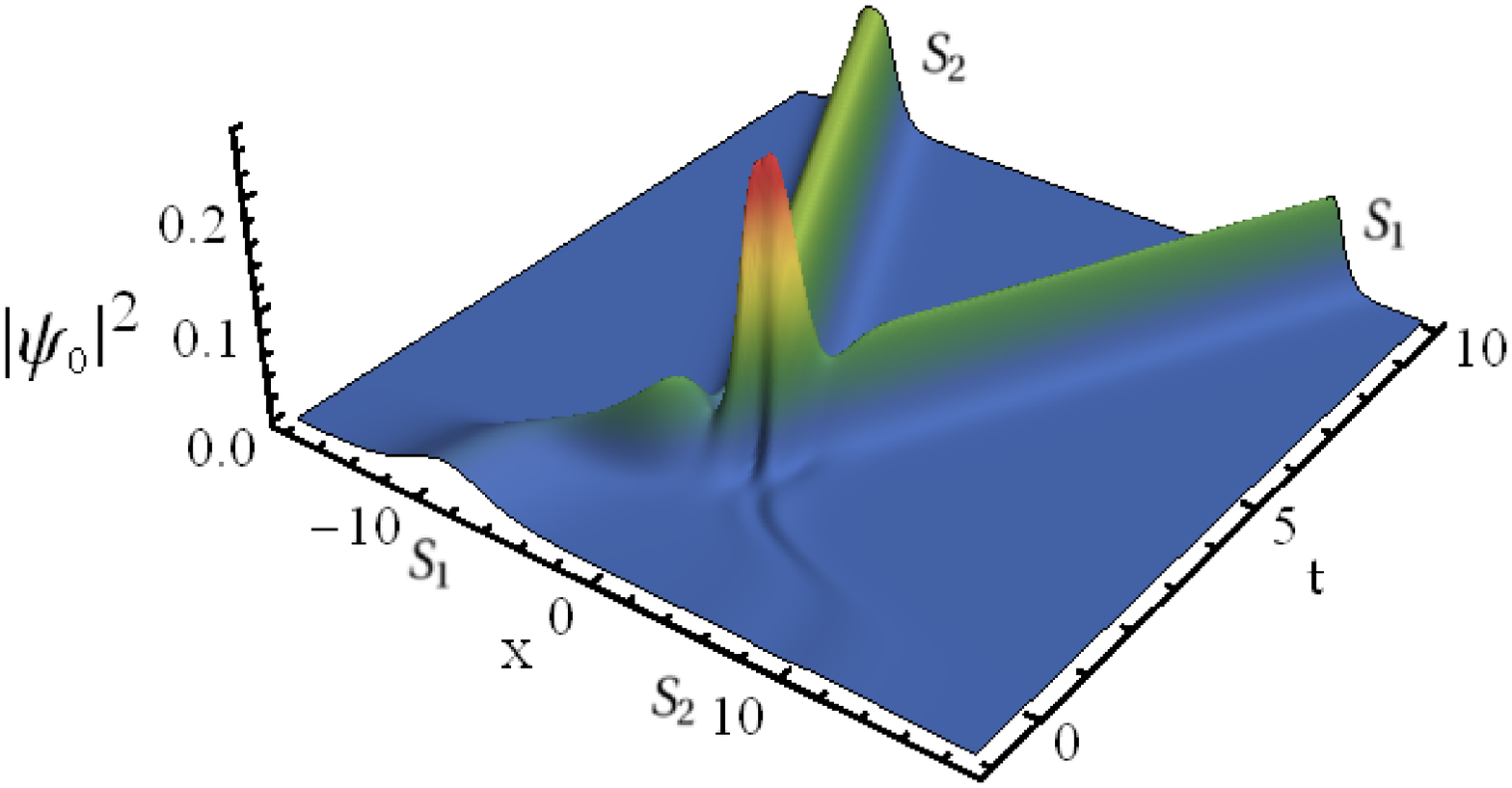}
\caption{Spin-switching interaction of PS with FS in autonomous 3-GP system (\ref{ccnls}) (top panels) and in non-autonomous 3-GP system (\ref{ncgp}) with time-varying kink-like nonlinearity coefficient and inhomogeneous potential given by (\ref{pot}) (bottom panels). The soliton parameters are $k_1 = -1.2 + i$, $k_2 = 1.5 - i$, $\alpha_1^{(1)} = \alpha_1^{(2)} = \alpha_1^{(3)} = 0.02$, $\alpha_2^{(1)} = 0.049,~\alpha_2^{(2)} =0.014,~\alpha_2^{(3)} = 0.049$, $\xi_1 = 0.2$, $\xi_2 = 0$, $\delta = -0.2$, and $\omega = 2.5$.}
\label{intfspsfig}
\end{figure}

From the asymptotic analysis given in Appendix C, we find that the amplitudes of non-autonomous FS $(S_1)$ and polar $(S_2)$ solitons before and after interaction can be related as $A_j^{l+}=T_j^{(l)}~A_j^{l-},~l=1,2,~j=1,2,3$. Here the expressions for transition amplitudes $T_j^{(l)}$'s are obtained as
\bes\bea
T_j^{(1)}&=& \frac{(k_1-k_2)(k_1^*+k_2)}{(k_1^*-k_2^*)(k_1+k_2^*)},\\
T_j^{(2)}&=&\left(\frac{(k_1+k_2^*)\Omega}{(k_1^*+k_2)(k_1^*-k_2^*)^2\alpha_2^{(j)} \alpha_2^{(4-j)*}\kappa_{11}^2 \Gamma_2^*}\right)^{\frac{1}{2}},
\eea\label{tr3c}\ees
where $\Omega=(\alpha_1^{(j)}\Gamma_3^*+(k_1^*-k_2^*)(\alpha_2^{(4-j)*}\kappa_{11}-\alpha_1^{(4-j)*}\kappa_{12})) (\alpha_1^{(4-j)*}\Gamma_3+(k_1-k_2)(\alpha_2^{(j)}\kappa_{11}-\alpha_1^{(j)}\kappa_{21}))$. Note that for the FS $(S_1)$ $|T_j^{(1)}|=1$. Thus the densities of non-autonomous FS($S_1$) after interaction are same as that of before interaction in the three components and is not influenced by any spin-mixing effects. Hence it undergoes elastic interaction with non-autonomous PS($S_2$) but suffers a phase-shift given by $\Phi_1=\frac{\theta_{22}-\epsilon_{22}-R_1}{2k_{1R}}\equiv \frac{1}{k_{1R}}\ln\left(\frac{(k_1-k_2)(k_1^*-k_2^*)}{(k_1+k_2^*)(k_1^*+k_2)}\right) $. The non-autonomous PS($S_2$) exhibits spin-switching among the components and is strongly influenced by the spin polarization parameters (see Eq.~\ref{tr3c}b)). Here non-autonomous PS($S_2$) undergoes a novel type of spin-switching interaction and also it experiences a phase-shift $\Phi_2=\frac{\theta_{22}-\epsilon_{22}-R_1}{2k_{2R}}\equiv \frac{k_{1R}}{k_{2R}}\Phi_1$ which is caused due to the spin-mixing nonlinearity. In this case, the change in relative separation distance between the two interacting solitons is given by $\Delta t_{12}=\left(1-\frac{k_{1R}}{2k_{2R}} \right)\Phi_1$.

This type of soliton interaction also takes place in multicomponent coherently coupled NLS system arising in the context of nonlinear optics \cite{tkjpa}. The possibility of making use of this property advantageously in matter wave switches, in which FS is considered as switch and PS is taken as a signal, has been suggested in Ref. \cite{Ieda}. Thus the appropriate inhomogeneous potential and the time-varying nonlinearity satisfying (\ref{ccnls}d) retains the nature of autonomous soliton interaction in the non-autonomous system (\ref{ncgp}) also, but affects the soliton parameters according to their form.

In Fig.~\ref{intfspsfig}, we have shown the non-autonomous (bottom panel) soliton interaction between FS($S_1$) and PS($S_2$). The non-autonomous FS($S_1$) undergoes elastic collision along with modulation in its amplitude according to the kink-like nonlinearity coefficient. But PS($S_2$) having single-hump (double-hump) profile in $\psi_{\pm 1}$ ($\psi_0$) switches its density profile to a double-hump (single-hump) with suppression (enhancement) of number density. In the top panel we present the collision of soliton in autonomous system (\ref{ccnls}) for comparison. One can note that the soliton collision is slower in the non-autonomous case.

\subsection{\textbf{Interaction between non-autonomous PSs}}
The interaction among two non-autonomous polar solitons can be obtained for the choice $\Gamma_j\neq 0,~j=1,2,3,$ and the corresponding asymptotic expressions are given in Appendix C. From those asymptotic expressions, we can observe that the modular amplitudes of solitons after interaction are same as that of before interaction i.e., $|A_j^{l+}|=|A_j^{l-}|,~l=1,2$, and $j=1,2,3$. This reveals the fact that the non-autonomous PSs always undergo elastic interaction without change in their amplitudes after interaction. However, the interacting PSs exhibit a phase-shift of $\Phi_1=\frac{R_4-\epsilon_{11}-\epsilon_{22}}{4k_{1R}} \equiv \frac{1}{ k_{1R}}\ln \left(\frac{(k_1 - k_2)(k_1^* - k_2^*)}{(k_1^* + k_2)(k_1 + k_2^*)}\right)$ for PS($S_1$) and $\Phi_2=\left(\frac{k_{1R}}{k_{2R}}\right)\Phi_1$ for PS($S_2$). Also, the change in relative separation distance between the two non-autonomous polar solitons before ($t_{12}^-=\frac{k_{1R}\epsilon_{22}-k_{2R}\epsilon_{11}}{4k_{1R}k_{2R}} $) and after interaction ($t_{12}^+=\frac{k_{1R}(R_4-\epsilon_{11})-k_{2R}(R_4-\epsilon_{22})}{4k_{1R}k_{2R}} $) can be written as $\Delta t_{12}=\left(1-\frac{k_{1R}}{k_{2R}} \right)\Phi_1$.

\begin{figure}[h]
\centering\includegraphics[width=0.4\linewidth]{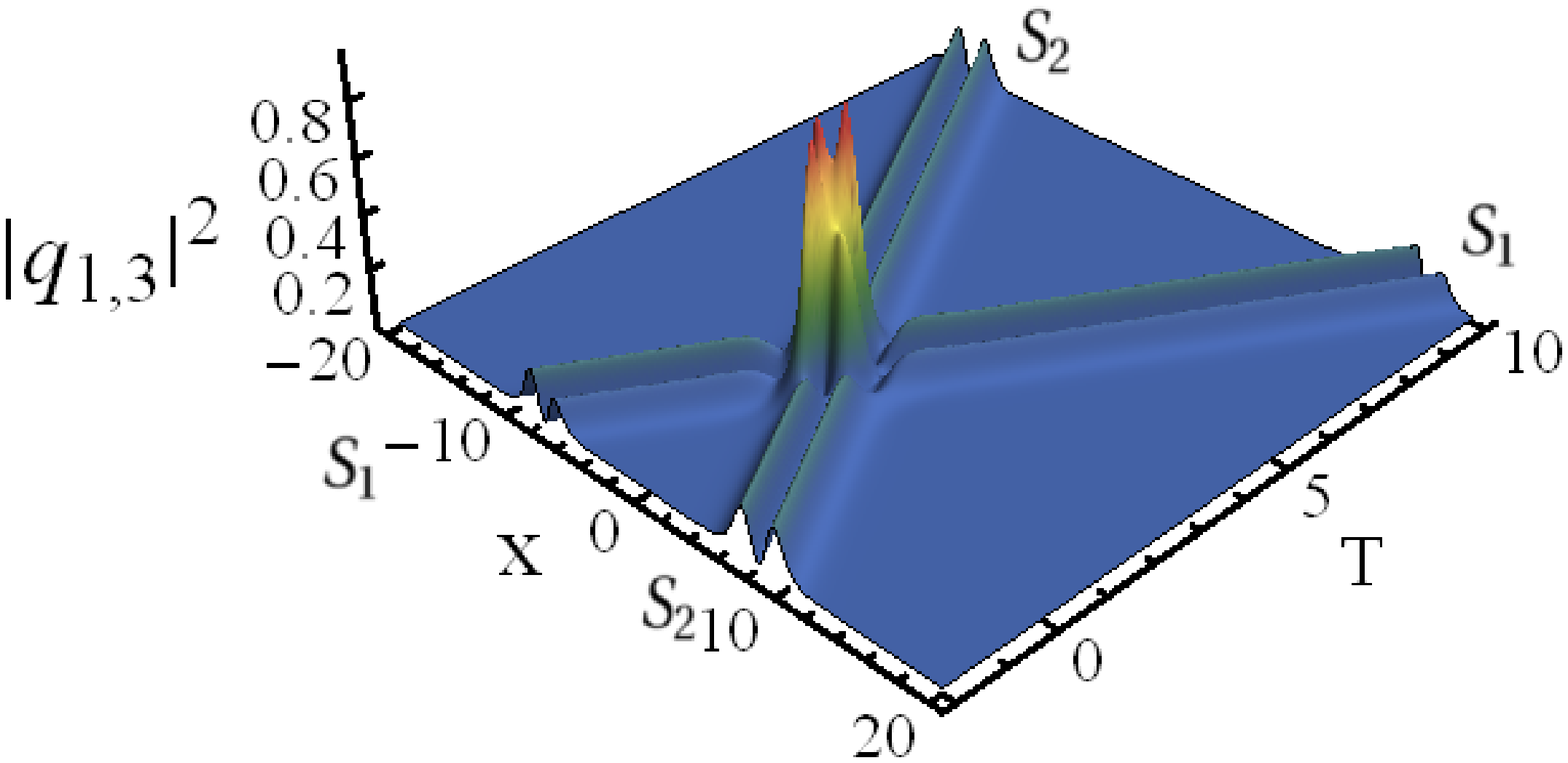}~~~~~~~\includegraphics[width=0.4\linewidth]{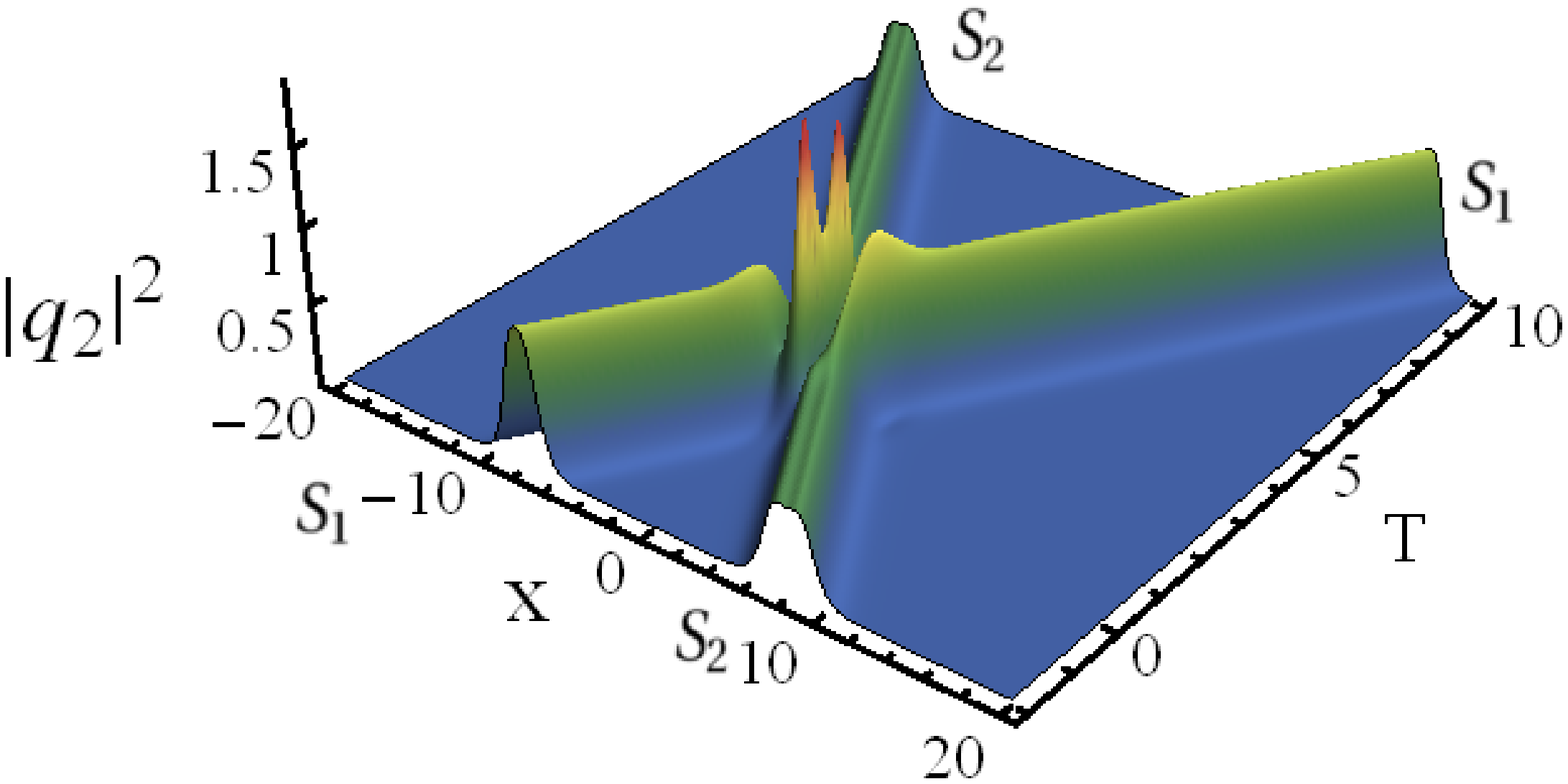}\\
\includegraphics[width=0.4\linewidth]{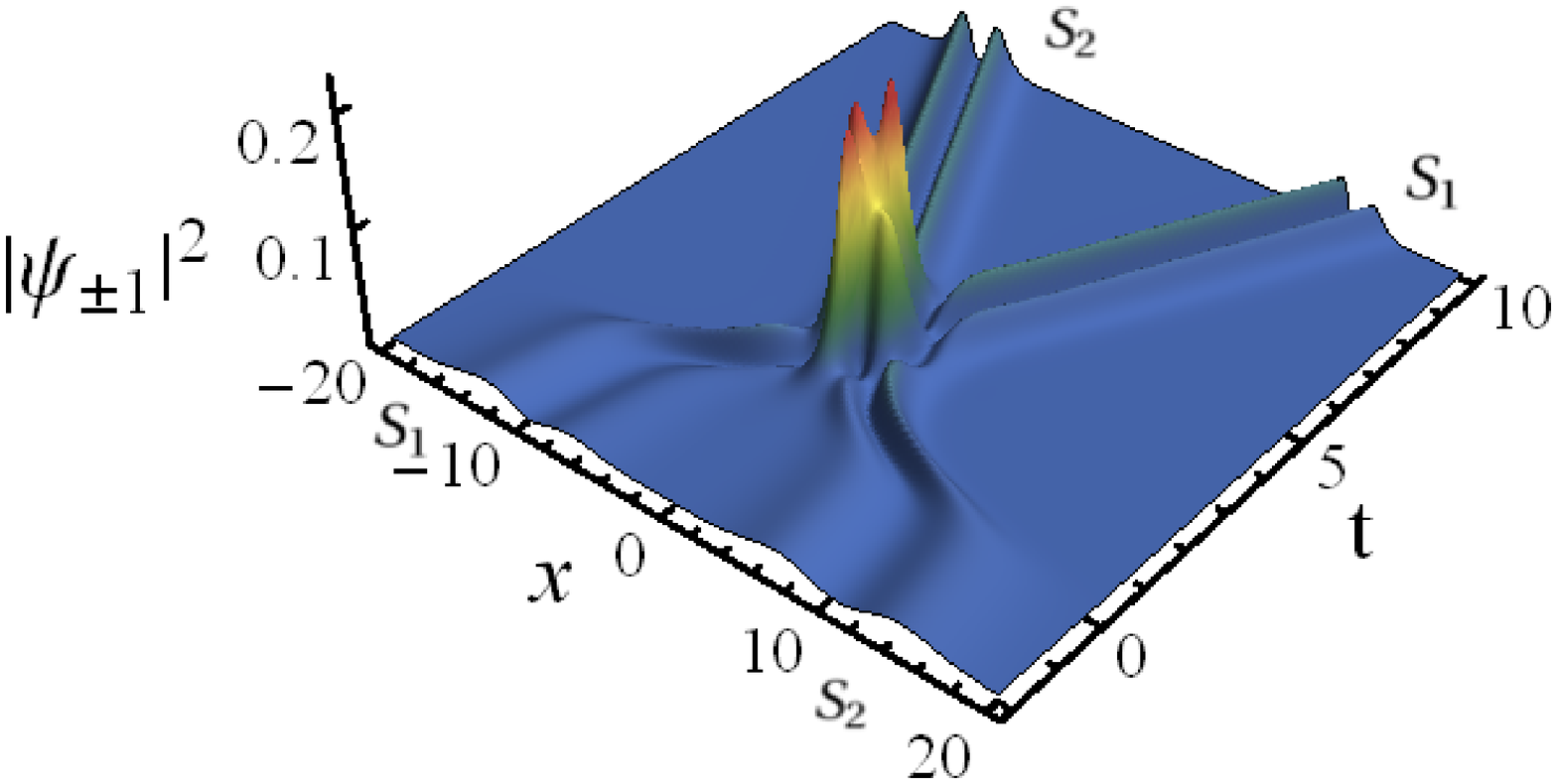}~~~~~~~\includegraphics[width=0.4\linewidth]{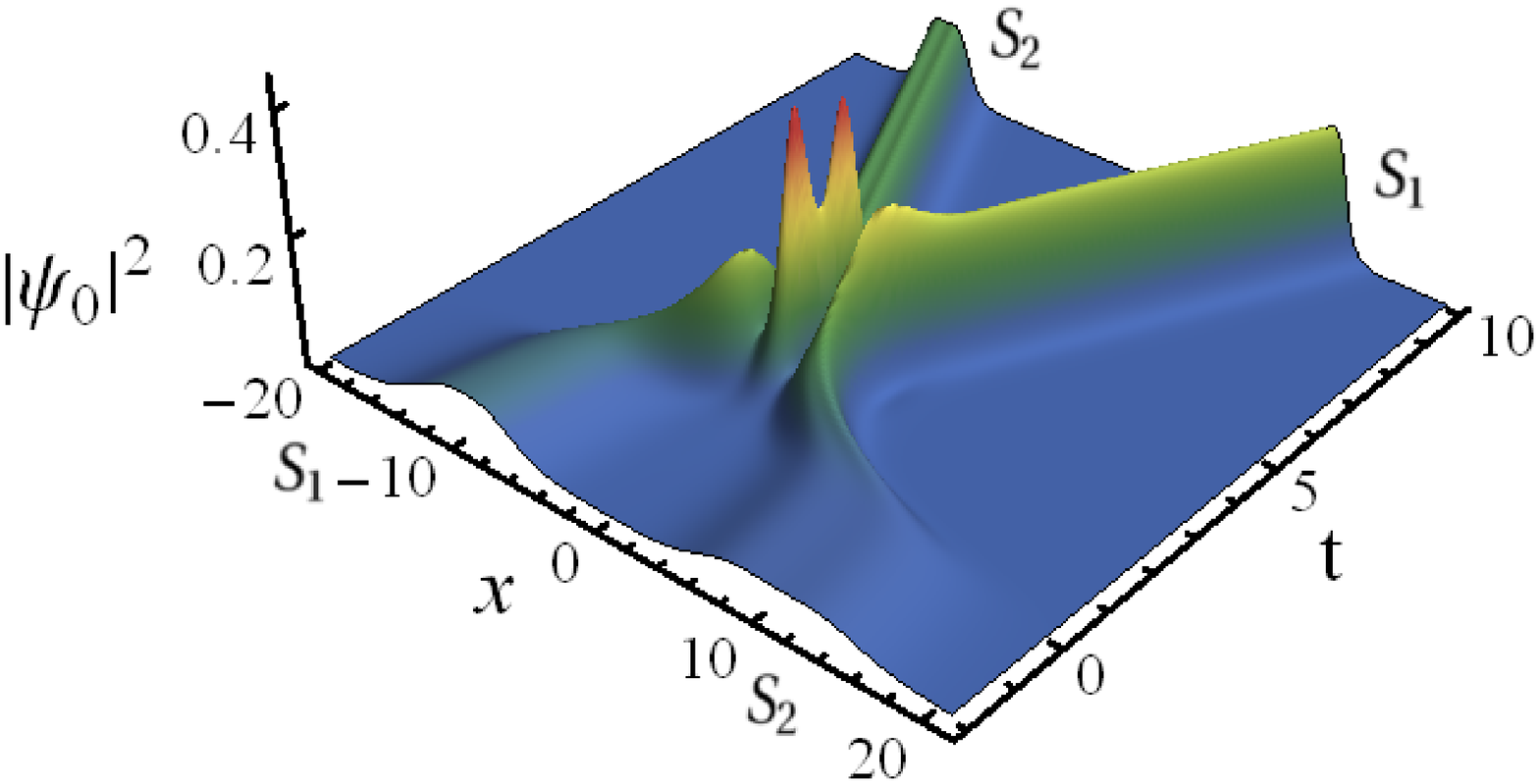}
\caption{Elastic interaction of two PSs in autonomous 3-GP system (\ref{ccnls}) (top panels) and in non-autonomous 3-GP system (\ref{ncgp}) with time-varying kink-like nonlinearity coefficient and inhomogeneous potential given by (\ref{pot}) (bottom panels).}
\label{int2psfig}
\end{figure}

Such type of elastic interaction of two non-autonomous PSs is shown in the bottom panels of Fig.~\ref{int2psfig} for $k_1=-1.2+i, ~k_2=1.1-i,~ \alpha_1^{(1)}= 0.020, ~\alpha_1^{(2)}=0.03,~\alpha_1^{(3)}=0.017, \alpha_2^{(1)}=0.03, \alpha_2^{(3)}=0.03,~ \alpha_2^{(2)}=0.04$, $\xi_1$ = 0.2, $\xi_2$ = 0, $\delta$ = -1, $\mbox{and}$ $\omega$ = 2. For completeness and comparison purpose, we have also given the interaction two PSs in the autonomous system (3) in the top panels of Fig.~\ref{int2psfig}.

\section{Summary}\label{conclusion}
In summary, we have transformed the non-autonomous three-coupled Gross-Pitaevskii equation (\ref{ncgp}) into a set of integrable autonomous 3-GP equations (\ref{ccnls}), along with a constraint in the form of the Ricatti equation, with the aid of a similarity transformation. In fact, the existence of non-autonomous matter wave solitons hinges on this constraint condition. First, we have obtained the exact soliton solutions of autonomous 3-GP equations by applying the Hirota's bilinearization method with a non-standard bilinearization procedure which involves an auxiliary function. Then by making use of these soliton solutions, the explicit soliton solutions of the non-autonomous GP system (\ref{ncgp}) are constructed.

We have obtained the explicit expressions for non-autonomous bright matter wave solitons for the kink-like nonlinearity and for the strength of the potential having the form (\ref{pot}). Here, we have demonstrated an interesting  compression of soliton accompanied by an amplification in the condensate density. We have also shown that such modulation form for time-dependent potential can very well be realized experimentally due to its close resemblance with the Hermite-Gaussian type function. Thus our study will have immediate applications in the context of soliton compression in spinor BECs as well as in multi-species BECs. We have shown that the non-autonomous matter wave solitons also admit different types of shape profiles namely, single-hump, double-hump and flat-top structures as in the case of autonomous bright solitons. However, now the profiles have been found to be modulated according to the chosen form of nonlinearity coefficient and inhomogeneous potential. Their interaction dynamics has been analyzed in detail and the effect of time-varying nonlinearity coefficient and inhomogeneous external potential have been studied. For this purpose, first we have revisited the collision dynamics of autonomous solitons in system (\ref{ccnls}) briefly. Particularly, our analysis has revealed the possibility of standard elastic collision of ferromagnetic solitons in system (\ref{ccnls}) for the choice $\Gamma_1=\Gamma_2=\Gamma_3=0$, in addition to the already reported non-trivial interaction between FSs leading to spin precession \cite{Ieda}. Thus, in the present work we have identified the parametric restriction for which spin precession (spin rotation) can not occur during the collision of FSs.

Then we have examined the effects of nonlinearity and the corresponding potential on soliton collisions. We have noticed that the non-autonomous bright matter wave solitons in system (\ref{ncgp}) undergo four types of interactions, namely (i) elastic interaction (without spin precession) between two FSs, (ii) spin precession interaction between two FSs, (iii) interaction with spin-switching in PS leaving FS unaffected and (iv) elastic interaction among the PSs, similar to the autonomous solitons. In all the above non-autonomous soliton interactions, it has been shown that there occurs a phase-shift and the role of temporal nonlinearity is to modulate the amplitudes of the solitons as well as the central position and hence the relative separation distance. We have also noticed the interesting point that in spite of the significant alteration in soliton parameters, the switching efficiency for the collision among non-autonomous FSs as well as collision nature between PS and FS remains to be the same as that of the autonomous solitons in system (\ref{ccnls}). This will lead to realize such non-trivial spinor soliton collisions for a wide range of nonlinearities and trapping potentials. Another important observation is that, for the kink-like nonlinearity, the non-autonomous matter wave solitons undergo slower collision than the autonomous matter wave solitons.

Thus, our theoretical analysis provides novel possibilities for controllable creation of bright solitonic matter waves and their compression in spinor BECs for kink-like nonlinearity modulation with suitable modulation in potential strength. Also, the present study can be straightforwardly extended to other types of temporal modulation of nonlinearities by finding appropriate strength of the potential from Eq.  (\ref{ccnls}d) or vice-versa. One can also introduce the gain/loss term in Eqs.  (\ref{ncgp}) and can very well extend the present analysis. Additionally, the exact non-autonomous bright soliton solutions reported in this work can serve as proper initial values for a direct numerical simulation of the general non-autonomous multicomponent GP equations (1). It is of future interest to investigate three-component spinor condensates in the presence of spatial and spatio-temporal inhomogeneities. We believe that this study will find important ramifications in the experiments on spinor condensates, matter wave switches and also in atom optics.

\section*{Acknowledgments}
The work of T.K. is supported by Department of Science and Technology (DST), Government of India, in the form of a major research project. R.B.M. acknowledges the financial support from DST in the form of Project Assistant. K.S. is grateful to the support of Council of Scientific and Industrial Research, Government of India, with Senior Research Fellowship. The authors also thank the principal and management of Bishop Heber College for constant support and encouragement.

\section*{Appendix A: Expression for various quantities appearing in two-soliton solutions and asymptotic analysis}
The various quantities appearing in two-soliton solution (\ref{sol2s}) and in the asymptotic analysis are defined below:
\bea
e^{R_u}&=&\frac{\kappa _{uu}}{(k_u+k_u^*)},~~ e^{\delta _0}=\frac{ \kappa _{12}}{(k_1+k_2^*)},~~
 e^{\delta _0^*}=\frac{ \kappa _{21}}{(k_2+k_1^*)},~~
e^{\delta _{uv}^{(j)}}=\frac{(-1)^{j+1} \alpha_v^{(4-j)*} \Gamma_u}{(k_u+k_v^*)^2},~~\nonumber\\
e^{\delta _u^{(j)}}&=&\frac{(-1)^{j+1} \alpha _u^{(4-j)*} \Gamma_3+(k_1-k_2) (\alpha _1^{(j)} \kappa _{2u}-\alpha _2^{(j)} \kappa_{1u})}{(k_1+k_u^*) (k_2+k_u^*)},~~
e^{\epsilon _{uv}}=\frac{\Gamma_u \Gamma_v}{(k_u+k_v^*)^4},\nonumber\\
 e^{\lambda _{uv}}&=&\frac{(k_1-k_2)^2 \kappa_{uv} \Gamma_{3-u}}{(k_u+k_v^*)(k_{3-u}+k_v^*)^2},~~~~~
e^{\tau _u}=\frac{\Gamma_u \Gamma_3}{ (k_u+k_1^*)^2 (k_u+k_2^*)^2},\nonumber\\
 e^{\mu _{uv}^{(j)}}&=&\frac{(k_1-k_2)^2 \alpha _{3-u}^{(j)} \Gamma_u \Gamma_v^*}{(k_u+k_v^*)^4 (k_{3-u}+k_v^*)^2},~~~~
e^{\theta _{uv}}=\frac{|k_1-k_2|^4 }{{\tilde{D}}(k_u+k_v^*)^2} \Gamma_u \Gamma_v^* \kappa_{3-u3-v},~\nonumber\\
 e^{\lambda _u}&=&\frac{(k_1-k_2)^4 ~\Gamma_1 \Gamma_2 \Gamma_u^*}{(k_1+k_u^*)^4 (k_2+k_u^*)^4},~~~~~~~
e^{\lambda _3}=\frac{(k_1-k_2)^4 \Gamma_1 \Gamma_2 \Gamma_3}{\tilde{D}},\nonumber\\
 e^{\phi_u^{(j)}}&=&(-1)^{(j+1)}{\alpha _{3-u}^{(4-j)*}}\frac{(k_1-k_2)^4 (k_1^*-k_2^*)^2~ \Gamma_1 \Gamma_2 \Gamma_u^*} {\tilde{D} {(k_1+k_u^*)^2(k_2+k_u^*)^2}},\nonumber\\
 e^{R_3}&=& \frac{|k_1-k_2|^2(\kappa_{11}\kappa_{22}-\kappa_{12}\kappa_{21})+|\Gamma_3|^2}{(k_1+k_1^*)|k_1+k_2^*|^2(k_2+k_2^*)}, ~~
e^{R_4}=\frac{|k_1-k_2|^8 |\Gamma_1|^2 |\Gamma_2|^2}{\tilde{D}^2},~~~~~\nonumber\\
 e^{\mu_u^{(j)}}&=&\frac{(k_1-k_2)^2~\Gamma_u}{\sqrt{\tilde{D}}(k_u+k_1^*)(k_u+k_2^*)}\left(\alpha_{3-u}^{(j)}\Gamma_3^*+(-1)^{(j+1)}(k_1^*-k_2^*)(\alpha_1^{(4-j)*}\kappa_{3-u2}-\alpha_2^{(4-j)*}\kappa_{3-u1})\right),\nonumber
\eea
where
$\Gamma_u= \alpha_u^{(1)}\alpha_u^{(3)}-(\alpha_u^{(2)})^2$, $\Gamma_3= \alpha_1^{(1)}\alpha_2^{(3)}+\alpha_2^{(1)}\alpha_1^{(3)}-2\alpha_1^{(2)}\alpha_2^{(2)}$, $\tilde{D}=(k_1+k_1^*)^2(k_1^*+k_2)^2 (k_1+k_2^*)^2 (k_2+k_2^*)^2$ and $\kappa_{uv}=\frac{ (\alpha_u^{(1)} \alpha_v^{(1)*}+2\alpha_u^{(2)} \alpha_v^{(2)*}+\alpha_u^{(3)} \alpha_v^{(3)*})}{(k_u+k_v^*)}$. Here $u,v=1,2$ and $j=1,2,3$.

\section*{Appendix B: Asymptotic analysis for two-soliton interaction of the autonomous 3-GP Eq.~(\ref{ccnls})}
Here, we present the results of the asymptotic analysis of two soliton solution of autonomous 3-GP equation (\ref{ccnls}), corresponding to the three broader types of matter wave soliton interactions given in Table 1. Without loss of generality, we consider $k_{1R}<0,~k_{2R}>0$ and $k_{1I}>k_{2I}$. Under this assumption for the two solitons $S_1$ and $S_2$ we find,
\bes\bea
 \mbox{Before interaction ($T \rightarrow -\infty$)} && S_1:~ \eta_{1R}\simeq0, ~~~ \eta_{2R}\rightarrow -\infty,~~~~~~~~~~~~~~~~~~~~~~~~~~~~~~~~~~~~\\
&& S_2: ~ \eta_{2R}\simeq0, ~~~ \eta_{1R}\rightarrow -\infty,\\
 \mbox{After interaction ($T \rightarrow +\infty$)~~} && S_1:~ \eta_{1R}\simeq0, ~~~ \eta_{2R}\rightarrow \infty,\\
&&  S_2: ~\eta_{2R}\simeq0, ~~~ \eta_{1R}\rightarrow \infty
\eea\label{asycon}\ees
In this section, the superscript (subscript) appearing in $q,~A,~P,~L$ represents the soliton (spin-component) number and the $-$ ($+$) sign appearing in the superscript denotes the soliton before (after) interaction.

\subsubsection*{\textbf{(i) Interaction of two FSs:}}
The ferromagnetic solitons having non-zero spin in the system (\ref{ccnls}) appear for the choice $\Gamma_j\equiv \alpha_j^{(1)} \alpha_j^{(3)}-(\alpha_j^{(2)})^2 = 0,~j=1,2$. Asymptotic expressions for FS($S_1$) and FS($S_2$) are given below.\\
\noindent{\textbf{Before interaction}}
\bes\bea
 \mbox{FS($S_1$):} \quad q_j^{1-} &=& A_j^{1-} \mbox{sech}\left(\eta_{1R}+\frac{R_1}{2}\right) e^{i\eta_{1I}},~ j=1,2,3,\\
 \mbox{FS($S_2$):}  \quad q_j^{2-} &=& A_j^{2-} \mbox{sech}\left(\eta_{2R}+\frac{R_2}{2}\right) e^{i\eta_{2I}}, ~ j=1,2,3,
\eea
where $A_j^{1-}=\frac{1}{2}\alpha_1^{(j)} e^{-\frac{R_1}{2}}$ and $A_j^{2-}=\frac{1}{2}\alpha_2^{(j)} e^{-\frac{R_2}{2}}$.\\
\noindent{\textbf{After interaction}}
\bea
 \mbox{FS($S_1$):} \quad  q_j^{1+} &=& A_j^{1+} \mbox{sech}\left(\eta_{1R}+\frac{R_3-R_2}{2}\right) e^{i\eta_{1I}},~ j=1,2,3,\\
 \mbox{FS($S_2$):} \quad  q_j^{2+} &=& A_j^{2+} \mbox{sech}\left(\eta_{2R}+\frac{R_3-R_1}{2}\right) e^{i\eta_{2I}},~ j=1,2,3,
\eea\label{intfs}\ees
where $A_j^{1+}=\frac{1}{2}e^{\delta_2^{(j)}-(\frac{R_2+R_3}{2})}$ and $A_j^{2+}=\frac{1}{2}e^{\delta_1^{(j)}-(\frac{R_1+R_3}{2})}$.  All the other quantities in Eqs.  (\ref{intfs}) are defined in Appendix A.

\subsubsection*{\textbf{(ii) Interaction of FS with PS:}}
Interaction between FS($S_1$) and PS($S_2$) can be achieved by choosing the soliton parameters to satisfy the conditions $\Gamma_1=0$ and $\Gamma_2\neq 0$, respectively, and their asymptotic expressions are given below.\\
\noindent{{\bf Before interaction}}
\bes\bea
 \mbox{FS($S_1$):~} && q_j^{1-}= {A_j^{1-}} \mbox{sech}\left(\eta_{1R}+\frac{R_1}{2}\right)e^{i\eta_{1I}}, \quad j=1,2,3,~~~~~~~~~~~~~~~~~~~~~~~~~\\
 \mbox{PS($S_2$):~} && q_j^{2-}={A_j^{2-}}\left(\frac{\mbox{cos}(P_j^{2-})\mbox{cosh}(\eta_{2R}^-)+i~\mbox{sin}(P_j^{2-})\mbox{sinh}(\eta_{2R}^-)}{4\mbox{cosh}^2(\eta_{2R}^-)+L^{2-}}\right) e^{i\eta_{2I}}, \quad j=1,2,3,~~~~~
\eea
where $A_j^{1-}=\frac{1}{2}\alpha_1^{(j)}e^{-\frac{R_1}{2}}$, $A_j^{2-}= 2e^{\frac{\delta_{22}^{(j)}+l_j^- -\epsilon_{22}}{2}}$, $P_j^{2-}= {\frac{\delta_{22I}^{(j)}-l_{jI}^-}{2}}$, $e^{l_j^-}=\alpha_2^{(j)}$, $L^{2-}=e^{(R_2-\frac{\epsilon_{22}}{2})}-2$, and $\eta_{2R}^-=\eta_{2R}+\frac{\epsilon_{22}}{4}$.\\
\noindent{{\bf After interaction}}
\bea
 \mbox{FS($S_1$):~} && q_j^{1+}={A_j^{1+}} \mbox{sech}\left(\eta_{1R}+\frac{\theta_{22}-\epsilon_{22}}{2}\right)e^{i\eta_{1I}}, \quad j=1,2,3,~~~~~~~~~~~~~~~~~~~\\
 \mbox{PS($S_2$):~} && q_j^{2+}={A_j^{2+}} \left(\frac{\mbox{cos}(P_j^{2+})\mbox{cosh}(\eta_{2R}^+)+i~\mbox{sin}(P_j^{2+})\mbox{sinh}(\eta_{2R}^+)}{4\mbox{cosh}^2(\eta_{2R}^+)+L^{2+}}\right) e^{i\eta_{2I}},~~~j=1,2,3,~~~~~
\eea\label{intfsps}\ees
where $A_j^{1+}= \frac{1}{2}e^{(\mu_{22}^{(j)}-\frac{\theta_{22}+\epsilon_{22}}{2})}$, $A_j^{2+}=2e^{\frac{\mu_{2}^{(j)}+\delta_1^{(j)}-\theta_{22}-R_1}{2}}$, $P_j^{2+}= {\frac{\mu_{2I}^{(j)}-\delta_{1I}^{(j)}}{2}}$, $L^{2+}=e^{R_3-(\frac{\theta_{22}+R_1}{2})}-2$, and $\eta_{2R}^+=\eta_{2R}+\frac{\theta_{22}-R_1}{4}$.

\subsubsection*{\textbf{(iii) Interaction of two PSs:}}
The detailed  asymptotic expressions for the interaction of two polar solitons of system (\ref{ccnls}) resulting for the general case $\Gamma_j \neq 0,~j=1,2,3,$ can be written as below. \\
\noindent{{\bf Before interaction}}
\bes\bea
 \mbox{PS($S_1$):}~q_j^{1-}={2A_j^{1-}} \left(\frac{\mbox{cos}(P_j^{1-})\mbox{cosh}(\eta_{1R}^-)+i~\mbox{sin}(P_j^{1-})\mbox{sinh}(\eta_{1R}^-)}{ 4\mbox{cosh}^2(\eta_{1R}^-)+L^{1-}}\right) e^{i\eta_{1I}},~j=1,2,3,~\\
 \mbox{PS($S_2$):}~q_j^{2-}={2A_j^{2-}} \left(\frac{\mbox{cos}(P_j^{2-})\mbox{cosh}(\eta_{2R}^-)+i~\mbox{sin}(P_j^{2-})\mbox{sinh}(\eta_{2R}^-)}{ 4\mbox{cosh}^2(\eta_{2R}^-)+L^{2-}}\right) e^{i\eta_{2I}},~j=1,2,3,~
\eea
where $A_j^{1-}= e^{\frac{\delta_{11}^{(j)}+l_j^{1-}-\epsilon_{11}}{2}}$, $P_j^{1-}= {\frac{\delta_{11I}^{(j)}-l_{jI}^-}{2}}$, $e^{l_j^{1-}}=\alpha_1^{(j)}$, $L^{1-}=e^{(R_1-\frac{\epsilon_{11}}{2})}-2$, $\eta_{1R}^-=\eta_{1R}+\frac{\epsilon_{11}}{4}$,  $A_j^{2-}= e^{\frac{\delta_{22}^{(j)}+l_j^{2-}-\epsilon_{11}}{2}} $, $P_j^{2-}= {\frac{\delta_{22I}^{(j)}-\l_{jI}^{2-}}{2}}$, $e^{l_j^{2-}}=\alpha_2^{(j)}$, $L^{2-}=e^{(R_2-\frac{\epsilon_{22}}{2})}-2 $, and $\eta_{2R}^-=\eta_{2R}+\frac{\epsilon_{22}}{4}$.\\
\noindent{{\bf After interaction}}
\bea
 \mbox{PS($S_1$):}~q_j^{1+}={2A_j^{1+}} \left(\frac{\mbox{cos}(P_j^{1+})\mbox{cosh}(\eta_{1R}^+)+i~\mbox{sin}(P_j^{1+})\mbox{sinh}(\eta_{1R}^+)}{4\mbox{cosh}^2(\eta_{1R}^+)+L^{1+}}\right) e^{i\eta_{1I}},~j=1,2,3,~ \\
 \mbox{PS($S_2$):}~q_j^{2+}=2A_j^{2+} \left(\frac{\mbox{cos}(P_j^{2+})\mbox{cosh}(\eta_{2R}^+)+i~\mbox{sin}(P_j^{2+})\mbox{sinh}(\eta_{2R}^+)}{ 4\mbox{cosh}^2(\eta_{2R}^+)+L^{2+}}\right) e^{i\eta_{2I}},~j=1,2,3,~
\eea\label{asy2ps}\ees
where $A_j^{1+}= e^{\frac{\phi_2^{(j)}+\mu_{22}^{(j)}-R_4-\epsilon_{22}}{2}} \equiv \frac{(k_1-k_2)(k_1^*+k_2)}{(k_1^*-k_2^*)(k_1+k_2^*)}A_j^{1-}$, $P_j^{1+}= {\frac{\phi_{2I}^{(j)}-\mu_{22I}^{(j)}}{2}}$, $L^{1+}=e^{(\theta_{22}-\frac{R_4+\epsilon_{22}}{2})}-2$, $\eta_{1R}^+=\eta_{1R}+\frac{R_4-\epsilon_{22}}{4}$, $A_j^{2+} = e^{\frac{\phi_1^{(j)}+\mu_{11}^{(j)}-R_4-\epsilon_{11}}{2}} \equiv \frac{(k_1^*-k_2^*)(k_1^*+k_2)}{(k_1-k_2)(k_1+k_2^*)}A_j^{2-}$, $P_j^{2+}= {\frac{\phi_{1I}^{(j)}+\mu_{11I}^{(j)}}{2}}$, $L^{2+}=e^{\theta_{11}-(\frac{R_4+\epsilon_{11}}{2})}-2$, and $\eta_{2R}^+=\eta_{2R}+\frac{R_4-\epsilon_{11}}{4}$.

\section*{Appendix C: Asymptotic analysis for the interaction of two non-autonomous matter wave solitons of 3-GP Eq.(\ref{ncgp})}
For the purpose of asymptotic analysis of Eq.~(\ref{ncgp}), we consider the same set of soliton parameter choices ($k_{1R}<0,~k_{2R}>0$ and $k_{1I}>k_{2I}$) as that of autonomous case. The resulting asymptotic behaviour of $\eta_{jR}$'s, $j=1,2$, are
\bes\bea
 \mbox{Before interaction ($t \rightarrow -\infty$)} && S_1:~ \widehat{\eta}_{1R}\simeq 0, ~~~ \widehat{\eta}_{2R}\rightarrow -\infty,~~~~~~~~~~~~~~~~~~~~~~~~~~~~~~~~~~~~\\
&& S_2: ~ \widehat{\eta}_{2R}\simeq 0, ~~~\widehat{\eta}_{1R}\rightarrow -\infty,\\
 \mbox{After interaction ($t \rightarrow +\infty$)~~} && S_1:~ \widehat{\eta}_{1R}\simeq0, ~~~ \widehat{\eta}_{2R}\rightarrow \infty,\\
&&  S_2: ~\widehat{\eta}_{2R}\simeq0, ~~~\widehat{\eta}_{1R}\rightarrow \infty.
\eea\ees
This is similar to the autonomous case (\ref{asycon}) and hence the asymptotic expressions for non-autonomous matter wave soliton in the non-autonomous 3-GP system(\ref{ncgp}) takes similar forms as that of the bright matter wave solitons in the integrable 3-GP equations (\ref{ccnls}), given by Eqs. (\ref{intfs}a)-(\ref{asy2ps}d) in the Appendix B, with the redefinition of the following quantities.
\bea
A_j^{l\pm} =  \widehat{A}_j^{l\pm}, \qquad
\eta_{lR}=\widehat{\eta}_{lR}, \qquad \eta_{lI}=\widehat{\eta}_{lI}+\tilde{\theta}, \quad l= 1,2, \quad j =1,2,3,
\eea
where $\widehat{A}_j^{l\pm} = A_j^{l\pm}\xi_1 \sqrt{2+\mbox{tanh}(\omega t +\delta)}$, $\widehat{\eta}_{lR} = k_{lR}\xi_1[2+\mbox{tanh}(\omega t +\delta)]x - [5t+\frac{1}{\omega}(4\ln[\mbox{cosh}(\omega t +\delta)]-\mbox{tanh}(\omega t +\delta))]2k_{lR}\xi_1^2(\sqrt{2}\xi_1 \xi_2+k_{lI})$, $\widehat{\eta}_{lI} = k_{lI}\xi_1[2+\mbox{tanh}(\omega t +\delta)]x -[5t+\frac{1}{\omega}(4\ln[\mbox{cosh}(\omega t +\delta)]-\mbox{tanh}(\omega t +\delta))]\xi_1^2(2\sqrt{2}\xi_1\xi_2 k_{lI}-k_{lR}^2 + k_{lI}^2)$, $l=1,2$, and $\tilde{\theta} = \left(\frac{-\omega \mbox{sech}^2(\omega t +\delta)}{2[2+\mbox{tanh}(\omega t +\delta)]}\right)x^2 + 2\xi_1^2\xi_2[2+\mbox{tanh}(\omega t +\delta)]x - 4\xi_2^2\xi_1^4 [5t+\frac{1}{\omega}(4\ln[\mbox{cosh}(\omega t +\delta)]-\mbox{tanh}(\omega t +\delta))]$.


\end{document}